\documentclass[copyright,creativecommons]{eptcs}
\usepackage{breakurl}
\usepackage[utf8]{inputenc}
\usepackage[T1]{fontenc}
\usepackage{todonotes}
\usepackage{amsmath}
\usepackage{amsthm}
\usepackage{amssymb}
\usepackage{stmaryrd}
\usepackage{bsymb,mybcode}
\usepackage{b2latex}
\usepackage{listings}
\usepackage{tikz}%[standalone]
\usepackage{environ}
\makeatletter
\newsavebox{\measure@tikzpicture}
\NewEnviron{scaletikzpicturetowidth}[1]{%
  \def\tikz@width{#1}%
  \begin{lrbox}{\measure@tikzpicture}%
  \BODY
  \end{lrbox}%
  \pgfmathparse{#1/\wd\measure@tikzpicture}%
  \BODY
}
\makeatother

\usetikzlibrary{arrows, automata, fit}
\usepackage{hyperref}
\usepackage{float}
\usepackage{array}
\usepackage{ragged2e}
\usepackage{mdframed}
\usepackage{xcolor}
\usepackage{setspace}

\usepackage{versions}
\includeversion{shortv}
\excludeversion{longv}

\newcommand\cmt[1]{}
\newcommand\fontb[1]{\textsf{#1}}

\definecolor{alexis}{RGB}{255,0,0}
\definecolor{hc}{RGB}{255,0,255}
\definecolor{dm}{RGB}{0,0,255}
\definecolor{lv}{RGB}{76,153,0} %% longversion

\newcommand{\textLV}[1]{{\textcolor{lv}{#1}}}

\lstdefinelanguage{DistAlgo}{
  language     = Python,
  morekeywords = {some, each, has, to, send, receive, msg, from_, at, deepcopy, update, setof, setup, run, start, sent, received, new, num, await}
}

\lstdefinestyle{mystyle}{
  mathescape,
  breaklines=true,
  columns=fullflexible,
  basicstyle = \ttfamily\upshape,
  escapechar = ~
}

\newcolumntype{C}[1]{>{\Centering\arraybackslash}p{#1}}
\newcolumntype{L}[1]{>{\raggedright\arraybackslash}p{#1}}
\newcolumntype{R}[1]{>{\raggedleft\arraybackslash}p{#1}}

\newcommand{\ie}[0]{\textit{i.e.}}
\newcommand{\eg}[0]{\textit{e.g.}}
\newcommand{\wrt}[0]{\textit{w.r.t.}}

\newcommand{\eqdef}{\,\mbox{\small$\stackrel{\mbox{\tiny $\triangle$}}{=}$}\,}

\newcommand{\context}[0]{\fontb{Context}}

\newcommand{\proc}[0]{proc}

\newcommand\axiom[1]{{\lbl{#1}}}

\newcommand\channel{\fontb{Channels}}
\newcommand\CHANNELSax{Channels}
\newcommand\send{\fontb{send}}
\newcommand\receive{\fontb{receive}} 
\newcommand\lose{\fontb{lose}}
\newcommand\sent{\fontb{sent}}
\newcommand\received{\fontb{received}}
\newcommand\intransition{\fontb{inChannel}}

\newcommand\NODES{\fontb{Nodes}}
\newcommand\NODESax{Nodes}
\newcommand\MESSAGES{\fontb{Messages}}

\newcommand\done{\fontb{done}}

\newcommand\neighboursN{network}
\newcommand\neighbours{\fontb{\neighboursN}}
\newcommand\STATES{\fontb{States}}
\newcommand\STATESax{States}

\newcommand\emptychannel{\fontb{emptyChannel}}

\newcommand{\sendingRequests}[0]{sr} 
\newcommand{\waitingAnswers}[0]{wa}  
\newcommand{\waitingRequests}[0]{wr}

\newcommand{\sendRequest}[0]{sendRequest}
\newcommand{\stopSending}[0]{stopSending}
\newcommand{\receiveAnswer}[0]{receiveAnswer}

\newcommand{\messPrefix}[0]{MessagePrefixes}
\newcommand{\enum}[1]{#1}

\newcommand\PCl{PCl}

\newcommand\chanvar{channels}

\newcommand{\tr}[2][]{\mathcal{T}_{#1}(#2)}
\newcommand{\trx}[2][]{\mathcal{T}^b_{#1}(#2)}
\newcommand{\tri}[2][]{\mathcal{G}^i_{#1}(#2)}
\newcommand{\trr}[2][]{\mathcal{G}^r_{#1}(#2)}
\newcommand{\tra}[2][]{\mathcal{A}_{#1}(#2)}
\newcommand{\trp}[2][]{\mathcal{T}^l_{#1}(#2)}
\newcommand{\XX}[0]{\overrightarrow{x}}
\newcommand{\YY}[0]{\overrightarrow{y}}
\newcommand{\bun}[0]{\cup}

\newcommand{\eventb}[0]{\textsc{Event-B}}
\newcommand{\distAlgo}[0]{\textsc{DistAlgo}}
\newcommand{\python}[0]{\textsc{Python}}
\newcommand{\rodin}[0]{\textsc{Rodin}}

\newcommand{\event}[1]{\emph{#1}}

\newtheorem{myexample}{\textbf{Example}}[section]
\newenvironment{example}
   {\begin{myexample}}
   {\end{myexample}}

\newtheorem{definition}{\textbf{Definition}}

\newcommand{\Cloc}[1]{\textsf{LC}(#1)}
\newcommand{\Vloc}[1]{\textsf{LV}(#1)}
\newcommand{\Evtloc}[1]{\textsf{Events}(#1)}

\newcommand{\grds}[1]{\textsf{Guards}(#1)}
\newcommand{\States}[1]{\textsf{StatesSet}(#1)}
\newcommand{\acts}[1]{\textsf{Actions}(#1)}

\newcommand{\Params}[1]{\textsf{Params}(#1)}

\floatstyle{boxed}
\restylefloat{figure}
\newcount\hour \newcount\minute
\hour=\time \divide \hour by 60
\minute=\time

\def\now{%
\ifnum \hour<13
  \ifnum \hour=0 \advance \hour by 12 \number\hour:\else \number\hour:\fi%
     \ifnum \minute<10 0\fi%
     \number\minute%
\ A.M.%
\else \advance \hour by -12 \number\hour:%
  \ifnum \minute<10 0\fi%
  \number\minute%
  \ P.M.%
\fi%
}

\title{Generating Distributed Programs from Event-B  Models}

\author{Horatiu Cirstea
\institute{LORIA UMR 7503\\ Universit\'e de Lorraine\\
  Vand\oe uvre-l\`es-Nancy, France}
\email{horatiu.cirstea@loria.fr}
\and Alexis Grall
\institute{LORIA UMR 7503\\ Universit\'e de Lorraine\\
  Vand\oe uvre-l\`es-Nancy, France}
\email{alexis.grall@loria.fr}
\and  Dominique M\'ery
\institute{LORIA UMR 7503\\ Telecom Nancy,  Universit\'e de Lorraine\\
  Vand\oe uvre-l\`es-Nancy, France}
\email{dominique.mery@loria.fr}}
\def\authorrunning{H. Cirstea, A. Grall \& D. Méry}
\def\titlerunning{Generating Distributed Programs from Event-B  Models}
\date{\today}

\begin{document}

\maketitle

%\tableofcontents

\begin{abstract}
  Distributed algorithms offer challenges in checking that they meet
  their specifications.  Verification techniques can be extended to
  deal with the verification of safety properties of distributed algorithms. In this paper, we
  present an approach for combining correct-by-construction approaches
  and transformations of formal models ({\eventb}) into programs
  ({\distAlgo}) to address the design of verified distributed
  programs. We define a subset LB (Local {\eventb}) of the {\eventb}
  modelling language restricted to events modelling the classical
  actions of distributed programs as internal or local computations,
  sending messages and receiving messages. We define then
  transformations of the various elements of the LB language into
  {\distAlgo} programs. The general methodology consists in starting
  from a statement of the problem to program and then progressively
  producing an LB model obtained after several refinement steps of the
  initial LB model.  The derivation of the LB model is not described in
  the current paper and has already been addressed in other works.
  The transformation of LB models into {\distAlgo} programs is
  illustrated through a simple example.  The refinement process and
  the soundness of the transformation allow one to produce
  correct-by-construction distributed programs.

\end{abstract}
%%%%%%%%%%%%%%%%%%%%%%%%%%%%%%%%%%%%%%%%%%%%%%%%%%%%%%%%%%%%%%%%%

%%% space before/after centred equations
\setlength\abovedisplayskip{2pt} %%
\setlength\belowdisplayskip{2pt}
%%% space before/after lstlisting
\setlength\medskipamount{2pt}

%%%%%%%%%%%%%%%%%%%%%%%%%%%%%%%%%%%%%%%%%%%%%%%%%%%%%%%%%%%%%%%%%
\section{Introduction}
%%============================================================================================
%%\subsection{Summary}
%%============================================================================================
{\eventb} is a formal modelling language developed by
Abrial~\cite{abrial2010} offering key features such as the use of set
theory as a data modelling notation, the use of refinement to
relate system models at different abstraction levels and the use of
mathematical proofs to verify consistency between refinement levels.
Moreover, the language is supported by the environment
{\rodin}\cite{abrial_rodin:_2010} which is extensible through the mechanism of
plugin.  Previous
works~\cite{ieee1394,DBLP:journals/ijaacs/TounsiMM16,DBLP:conf/isola/Mery18,DBLP:conf/ictac/Mery19}
illustrate the correct-by-construction design of distributed
algorithms using {\eventb} models and refinements; those works
show that at an adequate level of concretization of models, one can
derive a distributed algorithm in a pseudo algorithmic notation.
However, the derivation of concrete
{\eventb} models requires to develop a methodology related to a
given class of problems. For instance, we have produced a plugin
EB2RC~\cite{MeryM13,DBLP:conf/isola/ChengMM16} which automatically
generates a recursive algorithm from an {\eventb} model derived
by analysis of a problem such as Floyd's algorithm, or search
algorithms, or sorting algorithms.  The transformation of an
{\eventb} model into a recursive algorithm was based on the
definition of a class of (concrete) {\eventb} models satisfying
constraints making the transformation automatic.

\begin{center}
%\begin{center}
\begin{figure}[!tp] %[tb] 
\noindent 
\centering
\includegraphics{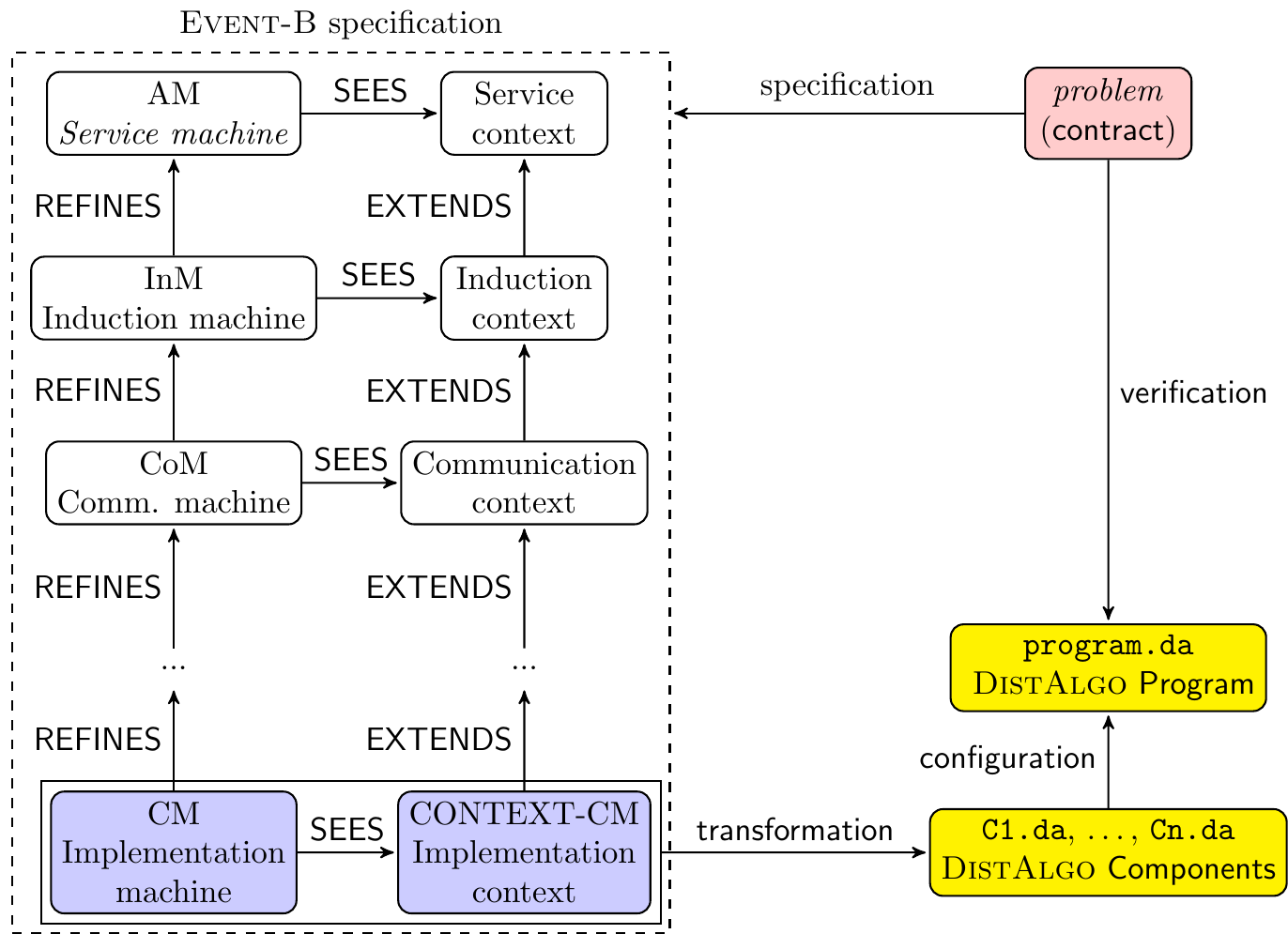}      
      
\caption{The global methodology for correct-by-construction distributed   algorithms.}
\label{fig:global}
\end{figure}
%\end{center}

\end{center}

In
the current paper, we study the systematic transformation of concrete
{\eventb} models
into
the \distAlgo~\cite{liu2012clarity}
programming language.
In fact, the design of a distributed algorithm using the
correct-by-construction approach
starts by expressing the
required computations in a very abstract {\eventb} model (AM)
and then progressively refining the model into a final concrete model
(CM) very close to an algorithmic expression of the distributed
algorithm. The main advantage of such a refinement-based process is
the preservation of safety properties of the different models: the
refinement is checked by discharging a list of proof obligations.  We
do not describe the process for developing the model CM which is
supposed to be a local {\eventb} model and which could be
translated into an algorithmic distributed notation. We focus on the
transformations required for obtaining a {\distAlgo} program from a
local {\eventb} model as indicated in Figure \ref{fig:global}: the
program \texttt{program.da} is generated from \textsc{CM} and
\textsc{CONTEXT-CM}. We will not provide the proof of correctness of
the translation but we will give enough details for trusting it. The
proof will be given in a future work.
\noindent 

~\\[-60pt]
\paragraph{An overview of the  integrated development framework}
%% \textbf{An overview of the  integrated development framework}
%%============================================================================================
%
Figure \ref{fig:global} provides an overview of our integrated
development framework for refinement-based program verification of
distributed algorithms.  The general methodology starts by stating the
\emph{problem} to solve by listing the requirements ({\ie} the
contract) attached to the problem; the requirements can be either
expressed in a formal language or in an informal textual language.
One has then to specify the {\eventb} machine \textsc{AM}
translating the main requirements for the given problem.  Then a list
of formal {\eventb} refined machines are produced to obtain a final
{\eventb} machine and context, \textsc{CM} and
\textsc{CM-CONTEXT}.  Finally, the translations of these final context
and machine into {\distAlgo} components and programs are generated
in two main steps: the automatic compilation of \textsc{CM} and
\textsc{CM-CONTEXT} into a {\distAlgo} program, and the
%% possible
manual tuning of the obtained {\distAlgo} components (if some
configurations were not specified in the model).

The refinement block (with nodes \textsc{AM}, \textsc{CONTEXT-AM},
\textsc{CM} and \textsc{CONTEXT-CM}) in Figure \ref{fig:global},
illustrates the mechanism for deriving machines via refinement.
%% LONG version
\begin{longv}
  \textLV{It can
be explained briefly as follows:
\begin{itemize}
\item  The machine \textsc{AM} defines events
  having the same contract as that expressed in the 
  requirements. This machine \textsc{SEES} the 
  \textsc{service context}, which expresses static information about the
  machine.  
\item The 
  machines \textsc{InM} and \textsc{CoM} are introducing respectively
  the description of the computing process and the corresponding
  communications.
\item The 
  machine \textsc{CM} refines \textsc{CoM}
  generating a concrete specification that satisfies the
  requirements. This machine \textsc{SEES} the 
  context \textsc{CONTEXT-CM}, which introduces control information
  for the new machine.
\item The labelled actions \textsc{REFINES}, \textsc{SEES} and
  \textsc{EXTENDS}, are supported by the {\rodin} platform and
  are 
  verified \textit{completely} using the proof assistant provided
  by {\rodin}.
\end{itemize}}  
\end{longv}
%% LONG version
The result of  the refinement is the {\eventb} machine
\textsc{CM}, which  contains  the  refined events   and  the proof
obligations  that  must  be discharged in   order to  prove   that the
refinement is correct.

Transformations of this {\eventb} machine \textsc{CM} into a
{\distAlgo} program is based on the extraction of information concerning the
network and the process classes   from the context \textsc{CONTEXT-CM},
and on the analysis of the localization of the different
variables. The events of \textsc{CM} are supposed to be  local  which
means that they are using only local instances of variables.
For instance, $pc$ will be a local variable with an instance $pc(p)$
for the process $p$. We will define precisely the localization process
from the code of {\eventb} models.
Finally, some constants whose values are not defined in the context
are instantiated during the configuration phase.

~\\[-40pt]
%%\subsection{Related work }
\paragraph{Related work }
\label{sec:RelatedWork}
%%============================================================================================
%

%% LONG version
\begin{longv}
  \textLV{From previous experience, we illustrate how the refinement can improve
and facilitate the verification process by relating state-based
models.}
\end{longv}
%% LONG version

We described a simple extension of the \textit{call-as-event}
paradigm~\cite{ijsi,MeryM13} to handle the design of concurrent
programs in the \textit{coordination}-based  approach but
we do not target a specific programming language as \distAlgo{}.   The
EB2ALL (http://eb2all.loria.fr) framework provides a list of transformations of
{\eventb} models into classical programming languages (C, C++, Java,
\ldots)  but it does not consider distributed algorithms. The current work can be considered as adding a new target
programming language but with the target of a distributed program like
it was proposed in Visidia (http://visidia.labri.fr) together with {\eventb}
with the plugin B2VISIDIA relating the local
{\eventb} model and a VISIDIA program. However, the VISIDIA approach
addresses distributed programs defined as set of rewriting rules of
graphs, which is less concrete and effective  than \distAlgo{} programs.
Code generation from classical
B models are supported by the Atelier~B  (http://www.atelierb.eu) tools but
those transformations do not consider  distributed programming models.
Atelier~B supports code generation into Ada, C, C++. Moreover, it is
defined over  Classical B software models restricted to the B0
language which is a computable subset of the B language but without
communications features. 
An EventB2Java~\cite{eventb2java} tool for {\rodin} has been
developed for translating any {\eventb} specification into
(sequential) JML or
Java code.  Finally, a  Tasking {\eventb}~\cite{edmunds2011tasking} for
{\rodin} extends the {\eventb} language to provide features for
specifying concurrent multi-tasking systems.  A model is decomposed
into several tasking machines which schedule and perform tasks
involving shared machines which correspond to protected resources
accessed by tasking machines.  The plugin provides a tool support for
translating a tasking specification into \textsc{Ada} code.  The
generated programs are not distributed ones  and consider only a
subclass of the ADA language.  Our work focuses on generating
{\distAlgo} programs from local  {\eventb} models and provides a way to
preserve powerfull safety properties from the local models.

~\\[-40pt]
%%\subsection{Overview of the paper}
\paragraph{Overview of the paper}
\label{sec:overview-paper}
%%============================================================================================
%
In the next section, we briefly present the two languages {\eventb}
and {\distAlgo}.
%% LONG version
\begin{longv}
  \textLV{We introduce also the modelling technique of
    G. Tel~\cite{tel2000introduction} for expressing distributed
    algorithms at the abstract level where a distributed algorithm is a
    set of local algorithms and each local algorithm is able to do either
    an internal action, or sending a message, or receiving a message.}  
\end{longv}
%% LONG version
%
Section~\ref{sec:modelling}  shows how distributed programs can be
modelled
in the sub-language called LB for Local \eventb.  Finally, in
Section~\ref{sec:translation} we define the transformation of LB
models into {\distAlgo} programs. Our paper then concludes with the
results and future work.
%% SHORT version
\begin{shortv}
  A more detailed description of the translation as well as the
  complete definition of the LB models and of the {\distAlgo} program
  of our example are available in~\cite{vpte2020:hal-02572971}.
\end{shortv}
%% SHORT version
%% LONG version
\begin{longv}
  \textLV{A more detailed description of the translation is given in
    Appendix~\ref{app:translation}. The complete definition of the
    machine \textsc{CM} and of context \textsc{CM-CONTEXT} used for
    the example translated in the paper is given in
    Appendix~\ref{app:mach}; the complete development of the model is
    given in Appendix~\ref{app:models}.The complete translation of the
    final context and machine into {\distAlgo} components and programs
    is given in Appendix~\ref{app:program}.  }
\end{longv}
%% LONG version

 %dm dm
\section{Modelling Distributed Programs}
\label{sec:modell-distr-progr}
%%============================================================================================
%% SHORT version
\begin{shortv}
  We describe briefly in this section the
  {\eventb} modelling language and the {\distAlgo}
  programming language.  
\end{shortv}
%% SHORT version
%% LONG version
\begin{longv}
  \textLV{We consider here the specification of distributed programs based on a
  model of computation due to G. Tel~\cite{tel2000introduction}.
  We describe briefly in this section this model as well the
  {\eventb} modelling language and the {\distAlgo}
  programming language.}
\end{longv}
%% LONG version
%
We will show later on how the corresponding specifications
are implemented following the methodology described in
Figure~\ref{fig:global}.

%% LONG version
\begin{longv}
  \subsection{\textLV{General Definitions  for Distributed Programs}}
  \label{sec:gener-defin-distr}
  %%============================================================================================
  Distributed algorithms and programs  can be expressed in  many
  programming languages (CSP, ADA, Java,  LINDA, \ldots)  and in many modelling
  languages (I/O automata, CCS, TLA$^+$, UNITY,\ldots).    In our
  approach, we intend to relate  one modelling language {\eventb} and one
  programming language {\distAlgo} through  the model of distributed
  computation due to G. Tel~\cite{tel2000introduction} 
  used often for describing basic and advanced distributed algorithms.
  
  \begin{definition}{(local and distributed algorithms~\cite{tel2000introduction})}
    %%---------------------------------------------------------------------------%
    Given a set ${\cal LC}$ of configurations, a set ${\cal
      LI}\subseteq{\cal LC}$ of initial configurations, and a set ${\cal
      M}$ of messages, a local algorithm ${\cal LA}$ is a structure
    $({\cal LC}, {\cal
      LI},\\\longrightarrow_i,\longrightarrow_s,\longrightarrow_r,{\cal
      M})$ with:
    \begin{itemize}
    \item $\longrightarrow_i\subseteq$ 
      ${\cal LC}\times{\cal LC}$ modelling internal computation steps,
    \item $\longrightarrow_s\subseteq$ 
      ${\cal LC}\times{\cal M}\times{\cal LC}$  modelling sending steps,
    \item $\longrightarrow_r\subseteq$ 
      ${\cal LC}\times{\cal  M}\times{\cal LC}$   modelling receiving
      steps.
    \end{itemize}
    
    A distributed algorithm for a collection of processes is a collection
    $\{{\cal LA}_1, \ldots, {\cal LA}_{n}\}$ of local algorithms, one
    algorithm ${\cal LA}_k=({\cal LC}_k, {\cal
      LI}_k,\longrightarrow_i^k,\longrightarrow_s^k,\longrightarrow_r^k,{\cal
      M})$ for each process $P_k$, with a transition relation
    $\longrightarrow$ defined over the set ${\cal C} = {\cal LC}_1\times
    \ldots \times {\cal LC}_n \times ({\cal M} \tfun \mathbb{N})$ of
    configurations.
    Given two configurations $C=(C_1,\ldots,C_n,M)$ and
    $C'=(C'_1,\ldots,C'_n,M')$, we have $C \longrightarrow C'$ iff\\ $\exists
    k \in \{1, \ldots,n\}: (\forall j \in 1..n: j
    \neq k : C_j=C'_j)$ and
    \begin{itemize}
    \item (internal transition)  
      $C_k \longrightarrow_i^k\  C'_k \ \wedge\  M'=M
      $
    \item {(send transition)
      $\exists m \in {\cal M}:
      \left\{\begin{array}{l} \forall o\in{\cal M}\backslash\{m\} : M'(o)=M(o)\\
      \wedge\ M'(m) = M(m)+1 
      \wedge\ (C_k,m,  C'_k) \in  \longrightarrow_s^k    \end{array}\right.$}
    \item {(receive transition) 
      $\exists m \in {\cal M}: M(m)\neq0:
      \left\{\begin{array}{l}
      \forall o\in{\cal M}\backslash\{m\} : M'(o)=M(o)\\
      \wedge\ M(m) = M'(m)+1 
      \wedge\ (C_k,m,C'_k) \in \longrightarrow_r^k
      \end{array}\right.$}
    \end{itemize}
  \end{definition}
  %%---------------------------------------------------------------------------%
  
  Following this approach for specifying distributed algorithms we have
  thus to define a collection of processes $P_1$, \ldots, $P_n$  with a local algorithm
  attached to each process.  We first briefly introduce {\eventb}
  and {\distAlgo} and then define the local algorithms using these
  formalisms.
\end{longv}
%% LONG version

\subsection{The Modelling Framework: {\eventb}}
\label{sec:eevent-b-modelling}
%%=======================================================================
%% LONG version
\begin{longv}
  \textLV{This section describes the modelling components of the {\eventb}
    language~\cite{abrial2010}.}
\end{longv}
%% LONG version
The {\eventb} language~\cite{abrial2010} contains two main
components, the \textit{context} which describes the static properties
of a system using \textit{carrier sets} $s$, \textit{constants} $c$,
\textit{axioms} $A(s,c)$ and \textit{theorems} $T_c(s,c)$, and
the \textit{machine} which  describes behavioural properties of a system
using \textit{variables} $v$, \textit{invariants} $I(s,c,v)$,
\textit{theorems} $T_m(s,c,v)$, \textit{variants} $V(s,c,v)$ and
\textit{events} $evt$.  A context can be extended by another context,
a machine can be refined by another machine and a machine can use
the \textit{sees} relation to include other contexts.

An {\eventb} machine defines
a set of \textit{state variables}
$Var$, taking their values in a set $Val$, and possibly modified by a
set of \textit{events} $Events$.
A set of invariants $I_i(s,c,v)$ contains typing information and
required safety properties that must be satisfied by the defined
system.  Each event $evt = \Bkeyword{any}\; x\; \Bkeyword{where}\;$ $
G_{evt}(s,c,v,x) \; \Bkeyword{then} \; v:|P_{evt}(s,c,v,x,v')\;
\Bkeyword{end}$ is composed of parameter(s) $x$, guard(s)
$G_{evt}(s,c,v,x)$ and action(s) $v:|P_{evt}(s,c,v,x,v')$.  Unprimed
variables refer to the state  variables before the event occurs and
primed variables refer to the  state variables after observation of the
event.  The \emph{before-after} predicate $BA(evt)(s,c,v,v')$ for
$evt$ is defined by $(\exists\ x\ \bcdot\ G_{evt}(s,c,v,x) \band
P_{evt}(s,c,v,x,v'))$.  A state $st$ of a machine is an element of the set
$St_{EB} = Var \tfun Val$.
The value of a variable $u\in Var$ in the
state $st$ is $st(u)$ and is denoted $st\llbracket u\rrbracket$.  The
notation $\llbracket .\rrbracket$ is extended to the list of variables
$v = (v_1, \ldots, v_n)$ by stating $st\llbracket (v_1, \ldots,
v_n)\rrbracket = (st(v_1), \ldots, st(v_n))$.  Finally, $\llbracket
.\rrbracket$ is extended to handle (arithmetical, boolean) expressions
by inductive definition: $st\llbracket exp(v)\rrbracket =
exp(st\llbracket v\rrbracket/v)$. 
%% LONG version
\begin{longv}
  \textLV{Since events are defined by
    expressions with free occurrence of unprimed and primed variables for v
    as v and v', we have to extend $\llbracket .\rrbracket$ as follows.}
\end{longv}
%% LONG version
For two states $st_1$, $st_2$ and an expression $exp(v,v')$ on primed
variables $v'$ and unprimed variables $v$, $st_1\llbracket exp(v,v')
\rrbracket st_2$ is defined by $exp(st_1\llbracket v\rrbracket/v,
st_2\llbracket v\rrbracket/v')$
the value of expression $exp$ where
unprimed variables are evaluated in state $st_1$ and primed variables
are evaluated in state $st_2$.  When an event $evt$ is observed
between two states $st_1$ and $st_2$, then $st_1\llbracket
BA(evt)(s,c,v,v')\rrbracket st_2$ 
%% SHORT version
\begin{shortv}
  holds.
\end{shortv}
%% SHORT version
%%
%% LONG version
\begin{longv}
  \textLV{holds and we write $st_1
    \xrightarrow{evt} st_2$.  We assume that only one event can be
    observed at any transition and a transition between two states is
    written as $st_1 \rightarrow st_2$ which is equivalent to
    $\exists\ evt\in Event \bcdot\ st_1\xrightarrow{evt} st_2$. }
\end{longv}
%% LONG version
In this
paper, we write deterministic \textit{actions} of the form $v \bcmeq
E(s,c,v,x)$ that are equivalent to $v :| v' = E(s,c,v,x)$. Using the
transition relation over the set of states, we can define state properties
as safety or invariance and traces properties.  

The {\eventb} modelling language supports the
\textit{correct-by-construction} approach to design an abstract model
and a series of refined models for developing any large and complex
system.
%
%% LONG version
\begin{longv}
  \textLV{Refinements, introduced by the REFINES clause, transform an
    abstract model into a more concrete version by modifying the state
    description.  A refinement allows modelling a system gradually by
    introducing safety properties at various refinement levels.  New
    variables and new events may be introduced in a new refinement
    level.  These refinements preserve the relation between the
    refining model and its corresponding refined concrete model, while
    introducing new events and variables to specify more concrete
    behavior of a system.  The defined abstract and concrete state
    variables are linked by introducing the \textit{gluing
      invariants}.  The generated proof obligations ensure that each
    abstract event is correctly refined by its concrete version.  }  
\end{longv}
%% LONG version
%
{\rodin}~\cite{abrial_rodin:_2010} is an integrated development
environment
%% (IDE)
for the {\eventb} modelling language based on Eclipse. It includes
project management, stepwise model development, proof assistance,
model checking, animation and automatic code generation.
%% LONG version
\begin{longv}
  \textLV{ Once an
    {\eventb} model is modelled and syntactically checked on the {\rodin}
    platform then a set of proof obligations (POs) is generated using the
    {\rodin} proof engine. {\eventb} supports different kinds of proof
    obligations, such as invariant preservation, non-deterministic action
    feasibility, guard strengthening in refinements, simulation, variant,
    well-definedness etc. More details related to the modelling language
    and proof obligations can be found in \cite{abrial2010}.}
\end{longv}
%% LONG version

\subsection{The {\distAlgo}  Distributed Programming Language}
\label{sec:dist-distr-progr}
%%============================================================================================

\distAlgo~\cite{liu2012clarity} is a programming language
%% built on top of {\python} and
used to develop distributed algorithms by providing high level
programming mechanisms such as communication primitives for the
exchange of messages between a set of processes.

A {\distAlgo} program is composed of several process classes managed
by a \texttt{main} module (see Example~\ref{ex:main}). 
A process class is made of a \lstinline[language=DistAlgo]?setup?
method which initializes the class attributes, a
\lstinline[language=DistAlgo]?run? method for carrying out the main
execution flow, several \lstinline[language=DistAlgo]?receive?
methods for handling the reception of messages and other user defined
methods that may be called by the \lstinline[language=DistAlgo]?run?
method.
For each process class \texttt{PC}, the \texttt{main} module uses a
statement of the form
\lstinline[language=DistAlgo]?pset=new(PC,num=n)?  to build the set
\texttt{pset} of \texttt{n} processes running the algorithm specified
for \texttt{PC}. The \lstinline[language=DistAlgo]?setup? method is
 called for the processes in each class and the
\lstinline[language=DistAlgo]?start? directive is eventually used to
trigger the \lstinline[language=DistAlgo]?run? method of all
processes.

\begin{figure}[!t]
\centering  
\includegraphics[scale=0.95]{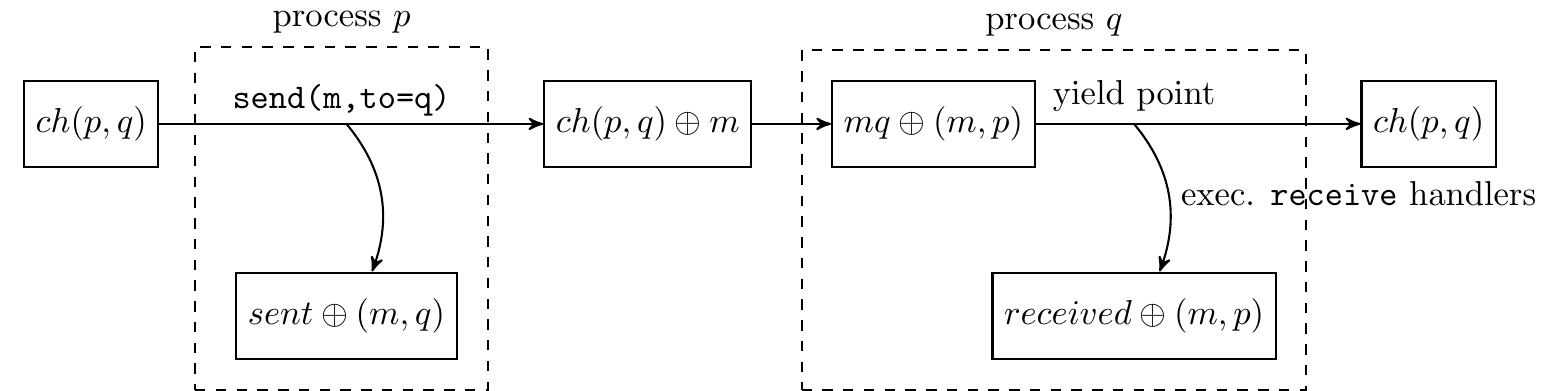}
  \caption{Comunications in {\distAlgo}: the communication channels
  $ch$, as well as the message queues $mq$ cannot be
  accessed explicitly in {\distAlgo}; only the sent and received
  messages can be accessed using the
  \lstinline[language=DistAlgo]?sent? and
  \lstinline[language=DistAlgo]?received? primitives.}
\label{fig:dacom}
\end{figure}

A process can send a \texttt{message} to another process \texttt{q}
with a statement \lstinline[language=DistAlgo]?send(message,to=q)?.
When a message arrives at the receiving process, it is put in a
message queue waiting to be received by the process.
To receive messages, the process control flow must be at a yield point
and this enables the receiving of every message in the message queue.
When a message is received, the \lstinline[language=DistAlgo]?receive?
message handlers
matching the message are
executed.  A yield point is a labeled statement
\lstinline[language=DistAlgo]?--l if await b1:s1 elif b2:s2 elif $\ldots$ elif bn:sn?
waiting for one of the conditions \lstinline[language=DistAlgo]?bi? to
hold in order to execute the corresponding branch
\lstinline[language=DistAlgo]?si?.
The history of sent and received messages can be accessed in
{\distAlgo} using the \lstinline[language=DistAlgo]?sent? and
\lstinline[language=DistAlgo]?received? primitives.
%% LONG version
\begin{longv}
  \textLV{The idea is that conditions on the
    history of sent and received messages can be used in the different
    methods of a local algorithm to determine the value of any
    variable.}
\end{longv}
%% LONG version
%
A graphical representation of the message exchanges is given in Figure~\ref{fig:dacom}.
Since {\distAlgo} is implemented as a {\python} module all the data
structures and primitives of the latter can be used. 
In our translation we use, in particular,
\lstinline[language=DistAlgo]?list?s, sometimes built using the
function \lstinline[language=DistAlgo]?range? which creates a list
interval of integers, and \lstinline[language=DistAlgo]?set?s, which
can be built from a list or using the function
%% \lstinline[language=DistAlgo]?setof?
\lstinline[language=DistAlgo]?setof($expr(x_1,\dots,x_n)$, $x_1$ in $S_1,\dots,x_n$ in $S_n$, $pred(x_1,\dots,x_n)$)?
which is a set comprehension with $expr$essions built out of elements
in the sets $S_1,\dots,S_n$ and satisfying a $pred$icate.
{\python} dictionaries are also
used; these can be updated
with the elements of another dictionary using the method
\lstinline[language=DistAlgo]?update? and cloned with the function
\lstinline[language=DistAlgo]?deepcopy? which copies an object and the
objects it contains recursively.
The {\distAlgo} boolean functions \lstinline[language=DistAlgo]?each?
and \lstinline[language=DistAlgo]?some? acting as a universal
quantifier and an existential quantifier respectively, are also used.

%% LONG version
\begin{longv}
  \textLV{We denote by $D^*$  the set of finite sequences with elements in $D$.
    A state of a {\distAlgo} program is an element of the set $St_{DA} =
    (Process\tfun Statement)\times(Process \tfun LocalState)\times(Process
    \tfun MsgQueue)\times ChannelStates$.  $Process$ is the set of
    processes and $Process\tfun Statement$ is a function which associates
    to each process the next statement to execute.  The local state of a
    process is an element of $Var \pfun Val$ associating $Val$ues to the
    $Var$iables.  $MsgQueue = (Val\times Process)^*$ is the set of message
    queues and $ChannelStates = Process\times Process \pfun Val^*$ is the
    set of possible states of the channels between each pair of processes.}

  \textLV{The operational semantics of {\distAlgo} is defined as a transition
    relation $\rightarrow$ between program states.  Given a function $f$,
    we write $f[x\rightarrow y]$ to specify that $f(x) = y$ and we denote
    by $f[x\bcmeq y]$ the function $f$ where $y$ is mapped to $x$. We
    present the transitions significant for this paper; a complete
    definition can be found in~\cite{liu2012clarity}.
    Given a process $p$, the assignment of a value $val$ to a variable $v$
    results in a change of the local state of $p$:  
    \[(P[p\rightarrow v \bcmeq val; s], ls[p \rightarrow ls_p], mq, ch)
    \rightarrow
    (P[p\bcmeq s], ls[p \bcmeq ls_p[v\bcmeq val]], mq, ch)\]
    When sending a message to a process $q$, we memorise it as
    sent and add it to the corresponding channel:
    \begin{align*}
      &(P[p\rightarrow send(m, to=q);s], ls[p \rightarrow ls_p], mq, ch[(p,q)\rightarrow ch_{p,q}]) \\
      &\rightarrow
      (P[p\bcmeq s], ls[p\bcmeq ls_p[sent\bcmeq ls_p(sent)@\langle(m,q)\rangle]], mq, ch[(p,q)\bcmeq ch_{p,q}@\langle m\rangle], mq)
    \end{align*}
    The arrival of a message to a process $q$ consists in adding it to the
    local message queue:
    \[
    (P, ls, mq[q\rightarrow mq_q], ch[(p,q)\rightarrow m.mt])\\
    \rightarrow
    (P, ls, mq[q\bcmeq mq_q@\langle (m, p)\rangle], ch[(p,q)\bcmeq mt])
    \]
    For the reception of a message at a label $l$ with a statement $s'$
    corresponding to the receive handler bodies we have:
    \begin{alignat*}{3}
      &(P[p&&\rightarrow l\ if\ await(b_1):s_1\ elif\ \ldots\ elif\ b_n:s_n;s], ls[p\rightarrow ls_p], mq[p\rightarrow m.mt], ch)\\
      &\rightarrow
      (&&P[p\bcmeq s';l\ if\ await(b_1):s_1\ elif\ \ldots\ elif\ b_n:s_n;s], ls[p\bcmeq ls_p[received \bcmeq ls_p(received)@\langle m\rangle]],\\
      &&&mq[p\bcmeq mt], ch)
    \end{alignat*}
    When all messages are received, {\ie} the queue is empty, and a condition
    $b_i$ of the await statement is satisfied we have:
    \begin{align*}
      (P[p\rightarrow l\ if\ await(b_1):s_1\ elif\ \ldots\ elif\ b_n:s_n;s], ls, mq[p\rightarrow \langle\rangle], ch)
      \rightarrow
      (P[p\bcmeq s_i;s], ls, mq, ch)
    \end{align*}
    %
    %% For unordered channels, a \emph{message reordering} transition
    %% is enabled and permutates the messages in any channel.
    %The program is executed on a single machine and is therefore
    %mainly useful as a simulation tool.  However, it is possible to
    %execute the process classes on multiple machines linked together
    %by modifying the main function for each machine.
}
\end{longv}
%% LONG version
%
%% %% SHORT version
%% \begin{shortv}
%% The operational semantics of {\distAlgo} is defined in~\cite{liu2012clarity} as a
%% transition relation over program states.
%% \end{shortv}
%% %% SHORT version

 %dm  dm
\section{Modelling Distributed Algorithms in {\eventb}}
\label{sec:modelling}
%%============================================================================================
We use the modelling technique of G. Tel~\cite{tel2000introduction}
and express a distributed algorithm as a set of local algorithms, each
local algorithm being able to do an internal action, or to send a
message, or to receive a message.  The final context \textsc{CONTEXT-CM} and
machine \textsc{CM} in Figure~\ref{fig:global} model such a distributed
algorithm using a subset of the modelling language {\eventb}, denoted
LB (Local {\eventb}).  We use the simple distributed algorithm
introduced in Example~\ref{ex:algoStar} to explain the methodology for
modelling algorithms following Tel's technique and the restrictions
imposed on
%% SHORT version
\begin{shortv}
  LB.
\end{shortv}
%% SHORT version
%% LONG version
\begin{longv}
  \textLV{ LB (to facilitate the translation towards \textsc{DistAlgo}).}
\end{longv}
%% LONG version

\begin{example}
\label{ex:algoStar}
%%---------------------------------------------------------------------------
We consider a distributed algorithm where each process $q$ in a set of
processes $Q$ sends its stored value to a central process $p$ who
previously made the corresponding requests.
%% LONG version
\begin{longv}
  \textLV{The communication channels
    between the requester and the other processes are reliable, {\ie}
    messages are neither lost nor modified, but the order in which
    messages are sent in a channel may change.}
\end{longv}
%% end LONG version
The local algorithm of the requester process $p$ has three states:\\
\begin{tikzpicture}[scale=0.5,->,>=stealth',shorten >=1pt,auto,node distance=4.5cm,thick, el/.style = {align=left, sloped}]
%% \begin{tikzpicture}[scale=0.3,->,>=stealth',shorten >=1pt,auto,node distance=4.5cm,thick, el/.style = {align=left, sloped}]
    \tikzstyle{every state}=[circle, draw, minimum size=1cm]

    %% \node[state, align=center, initial, initial text=INITIALISATION]  (sending) []   {$sending$\\$requests$}; 
    \node[state, align=center, initial, initial text=INITIALISATION]  (sending) []   {$\sendingRequests$}; 
    %% \node[state, align=center]  (waiting)  [right=3cm of sending]                      {$waiting$\\$answers$}; 
    \node[state, align=center]  (waiting)  [right of=sending]                      {$\waitingAnswers$}; 
    \node[state, align=center]  (done)         [right of=waiting]   {$\done$};

    \path   (sending) edge[loop above] node[at start, above left] {$\sendRequest$} (sending)
    edge node[align=center] {$\stopSending$} (waiting)   
    (waiting) edge[loop above] node[at start, above left] {$\receiveAnswer$} (waiting)
    %% edge node[align=center] {$p\_terminate$} (done)
    edge node[align=center] {$terminate$} (done)

            ;
\end{tikzpicture}
\\
While in state
%% $sending\_request$,
$\sendingRequests$, $p$ sends a request to each of the
processes in $Q$ and moves  to state
%% $\waitingAnswers$,
$\waitingAnswers$,
when all requests
have been sent.  In the state
%% $\waitingAnswers$,
$\waitingAnswers$,
it awaits for
answers from the processes in $Q$.  
When all answers are received,
process $p$ has terminated its local algorithm and moves to  state
{\done}.

\noindent
%% \begin{minipage}{1.0\linewidth}
  %% \begin{minipage}{0.65\linewidth}
    \begin{tikzpicture}[scale=0.5,->,>=stealth',shorten >=1pt,auto,node distance=4.5cm,thick, el/.style = {align=left, sloped}]
    \tikzstyle{every state}=[circle, draw, minimum size=1cm]

    %% \node[state, align=center, initial, initial text=INITIALISATION]  (waiting) []   {$waiting$\\$requests$}; 
    \node[state, align=center, initial, initial text=INITIALISATION]  (waiting) []   {$\waitingRequests$}; 
    \node[state, align=center]  (done)  [right of=waiting] {$\done$};

    \path   (waiting) edge[loop above] node[at start, above left] {$receiveRequestSendAnswer$} (waiting)
                    %% edge[] node[align=center] {$q\_terminate$}
                    edge[] node[align=center] {$terminate$}
                    (done)
            ;
\end{tikzpicture}
\\
  %% \end{minipage}
  %% \begin{minipage}{0.35\linewidth}
Each  process in $Q$ is initially in a state
%% $\waitingRequest$
$\waitingRequests$
in
which it waits for  a request from $p$ and moves to state {\done} after
receiving the request and sending its stored value. 
  %% \end{minipage}
%% \end{minipage}
%%---------------------------------------------------------------------------
\end{example}

The general architecture of the distributed algorithm (processes,
topology, channels, communications) is specified in the {\eventb}
context \textsc{CONTEXT-CM} while the list of events of the machine
\textsc{CM} induces the specifications of the local algorithms as
labelled transition systems.  In the sequel, the pair \textsc{CM} and
\textsc{CONTEXT-CM} defining the LB distributed model is called simply \textsc{CM}.

\subsection{Defining the General Architecture of the Distributed Program}
\label{sec:context}
%%============================================================================================
Sets, constants and corresponding axioms defined in the context of
a distributed model are of two categories: the general ones present in
the context of any algorithm and those which are specific to
the  modeled algorithms.
The most important elements of the context corresponding to the
algorithm described in Example~\ref{ex:algoStar} is given 
in Figure~\ref{fig:example}:
\begin{figure}[!t]
%\framebox{\makebox[0.9\width]{
%\resizebox{\textwidth}{!}{\boxed{
\centering  
  \begin{minipage}{0.9\linewidth}
      \CONTEXT{CONTEXT-CM}{C00}{}
\SETS{
	\Set{\NODES\ \STATES\ \MESSAGES}{// General sets}
	\Set{$\messPrefix$}{// Algorithm specific sets}
}
\CONSTANTS{
	\Constant{\neighbours}{// The topology (general)}
	\Constant{\channel\  \emptychannel\ \sent\ \received\ \intransition}{// Communication channels (general)}%%\ \readyForReception
	\Constant{\send\ \receive\ \lose}{// Communication primitives (general)}
  \Constant{$P$ $p$ $Q$}{// Process classes and processes (specific to the algorithm)}
	\Constant{$request$ $answer$}{// Algorithm specific constants}
	\Constant{$availableResources$}{// Algorithm specific constant}
  \Constant{$\sendingRequests{}$ $\waitingAnswers{}$ $\waitingRequests{}$ \done}{// Process states (specific to the algorithm except for \done, general)}
}
\AXIOMS{
	\Axiom{\NODESax}{false}{$partition(\NODES, P, Q)$}{// Partition of the set of processes}
	\Axiom{P}{false}{$partition(P, \{p\})$}{// Partition of the classes of processes}
	\Axiom{network\_typing}{false}{$\neighbours \in{} \NODES \tfun{} \pow{}(\NODES)$}{// Network specification}
	\Axiom{network\_value}{false}{$\neighbours = \{\proc\qdot{}\proc\in{}P|\proc\mapsto{}Q\}\bunion{}\{\proc\qdot{}\proc\in{}Q|\proc\mapsto{}\{p\}\}$}{}
	\BBcomment{States of the processes}
	\Axiom{\STATESax}{false}{$partition(\STATES, \{\sendingRequests\}, \{\waitingAnswers\}, \{\waitingRequests\}, \{done\})$}{}
  \BBcomment{Communication channels}
	\Axiom{\CHANNELSax}{false}{$\channel =
    \NODES\cprod{}\NODES\tfun{}(\MESSAGES\tfun{}\nat{}\cprod{}\nat{}\cprod{}\nat{})$}{}
	\BBcomment{Algorithm specific constants (types of exchanged messages, process resources)}
	\Axiom{\enum{\messPrefix}}{false}{$partition(\messPrefix,\{request\},\{answer\})$}{//@P@Q}
	\Axiom{availableResources\_typing}{false}{$availableResources \in{} Q \tfun{} \nat{}$}{}
	\BBcomment{Communication axioms (general to all algorithms)}
%	\item $\vdots$
}
\END
\end{minipage}
%}}
\caption{Sets and constants for the Example~\ref{ex:algoStar}}
\label{fig:example}
\end{figure}

For every distributed algorithm, the set {\NODES} of processes is
defined axiomatically as a partition into process classes, the
processes of each class featuring a similar local algorithm:
$$\axiom{\NODESax}:partition(\NODES, \PCl_1,\dots,\PCl_n)$$
For each process class $\PCl_i$ one can enumerate explicitly its
processes using an axiom
$$\axiom{\PCl}_i:partition(\PCl_i,\{\proc_1\},\dots,\{\proc_m\})$$
These partitions depend of course on the specific algorithm modeled
and, in general, the processes are not explicitly enumerated.
%
%% LONG version
\begin{longv}
  \textLV{
    For our example, we have the class $P$ of requester processes
    consisting of only one process $p$, and the class $Q$ of processes
    with stored values. The number and identities of these latter processes
    is not specified, keeping thus the model the most general possible at
    this stage.}
\end{longv}
%% LONG version

The topology, denoted $\neighbours$, is specified by a function
associating to each process its neighbours:
%
%% $\axiom{network\_typing}:$
$\neighbours\in{}\NODES\tfun{}\pow{}(\NODES).$
The concrete definition of the topology specific to the distributed
algorithm {under consideration} is specified using an axiom whose
general form should be
\begin{equation*}
  %% \begin{aligned}
  \begin{array}{r@{\hspace{3pt}}c@{\hspace{3pt}}l@{\hspace{3pt}}}
\axiom{network\_value}:\neighbours & = &  \{\proc\cdot \proc \in
\PCl_1|\proc\mapsto expr_1\} \bunion \\
&& \dots \bunion\{\proc\cdot \proc
\in \PCl_n|\proc\mapsto expr_n\}
  \end{array}
  %% \end{aligned}
  \end{equation*}
%% $$\axiom{network\_value}:\neighbours = \{\proc\cdot \proc \in
%% \PCl_1|\proc\mapsto expr_1\} \bunion
%% \dots \bunion\{\proc\cdot \proc
%% \in \PCl_n|\proc\mapsto expr_n\}$$
%
In the example, the topology is defined as a star with the process $p$
in the center.

As we will see later on, the \textit{control states} in {\STATES} are
used for structuring the observation of events in the local
algorithms.
%% The generation of local algorithms from the model \textsc{CM} uses
%% \textit{control states} for structuring the observation of events
%% simulating local algorithms.
The set of all possible control states of all processes is defined as
a partition by an axiom \axiom{\STATESax}. 
%% LONG version
\begin{longv}
  \textLV{The
    particular state $\done$ denotes the final state of any local
    algorithm.
    The explicit definition of the control states is used here to guide  the
    translation of the local algorithms  by grouping 
    events that are enabled in the same state.
    Explicit control states also enable the generation
    of state machines for visualizing the distributed algorithm.}
\end{longv}
%% LONG version
%
%% \cmt{An alternative approach would be the one used in
%% EB2ALL~\cite{mery2011automatic} to group the guards of the concrete
%% events which refine the same abstract events in order to optimize
%% the code.}

%% HC: To move in the discussions We only consider a static topology
%% in this paper; for dynamic one the topology should be specified as
%% a variable.

\begin{figure}[!t]
%%---------------------------------------------------------------------------
% \input{figures/lb_com}
\centering  
\includegraphics[scale=0.9]{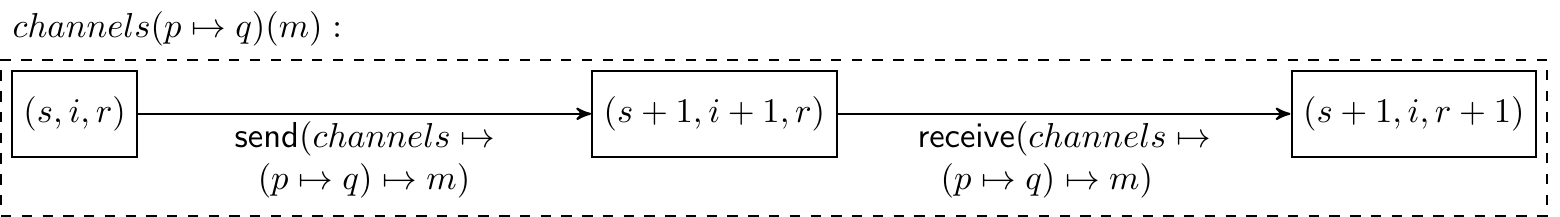}  
  \caption{We suppose that $m$ has been already sent, resp. received,
    $s$, resp. $r$ times, and that $i$ copies are in the channel:
    $\chanvar(p\mapsto q)(m) = (s,i,r)$. When we $send$ the message
    $m$ we increment the sent counter and the number of
    messages in the channel; when we $receive$ it we increment the
    received counter and decrement the number of messages in the channel.}
\label{fig:lbcom}
\end{figure}

The context
%% also defines
should also define a constant {\channel} modelling the set of
all possible values of communication channels between processes and
the set {\MESSAGES} of messages exchanged through these channels.
The current state of a channel between two
processes is defined as a multiset, corresponding to the messages that
were sent, received and in transition, {\ie} sent but not yet received
or lost. For instance, \sent(channel,p,q,mes) is returning how many
times the message mes has been sent by p to q. Hence, for each channel
we can retrieve the exchanged messages using the functions {\sent},
{\received} and {\intransition} of type
$\channel\cprod(\NODES\cprod\NODES)\cprod\MESSAGES\tfun\mathbb{N}$.
The functions {\send}, {\receive} and {\lose} of type
$\channel\cprod(\NODES\cprod\NODES)\cprod\MESSAGES\tfun\channel$
describe the transformation of a channel between two processes ({\ie}
adding or removing a message) when one operation (send, receive or
lose) is observed.
More precisely, we consider that the channels do not preserve the order
in which messages are sent, and {sending a message} consists in incrementing the
{\intransition} part and the {\sent} part of a channel between two
processes. 
%% LONG version
\begin{longv}
  \textLV{Checking if a message is ready to be received by a process
    from another for a value of the channels consists in checking that the
    message is in the {\intransition} part of the channel between the two
    processes.
The axioms specifying the behaviour and the characteristics of the
communication channels ({\eg} order preserving) as well as the
corresponding primitives are general, {\ie} do not depend on the
modeled distributed algorithm.}
\end{longv}
%% LONG version
%
The evolution of the channel between two processes $p$ and $q$
concerning a message $m$ is modeled in LB by the variable
$\chanvar(p\mapsto q)(m)$,
as depicted in Figure~\ref{fig:lbcom}. This variable
models the channel $ch$ and the message queues $mq$ as well as the
\lstinline[language=DistAlgo]?sent?  and
\lstinline[language=DistAlgo]?received? {\distAlgo} primitives
described in Section~\ref{sec:dist-distr-progr} (Figure~\ref{fig:dacom});
the transfer of the message from the channel to the message queue
which is builtin in {\distAlgo} is not explicitly modeled in LB.

%%---------------------------------------------------------------------------

Sets and constants mentioned above should be present in the
context of any distributed algorithm modeled in LB.  Other
enumerated sets defined necessarily as the disjoint union of
singletons using the $partition$ construct as well as
constants specific to the modeled algorithm can be defined in the
context.
The type of such a constant $cst$ is defined by an axiom of name
\axiom{cst\_typing} while its value may be defined
by an axiom of name \axiom{cst\_value}.
For instance, in our example we require that each process of $Q$ has a
non-negative integer $availableResources$.
For the purpose of our example, we also define an enumerated set
${\messPrefix}$ consisting of $request$ and $answer$ which
correspond to the two kinds of messages exchanged between the
processes.  
%% LONG version
\begin{longv}
  \textLV{For simplicity we consider here that any of these types is
    a sub-type of {\MESSAGES} and hide the injections we use in the real
    model.}
\end{longv}
%% LONG version

Annotations of the form $@\PCl_1\dots @\PCl_n$ are used to specify
that the annotated elements are \emph{local} to the processes in the
corresponding class.
{This is done either in the
  axiom \axiom{cst\_typing} to indicate the process classes concerned
  by the constant $cst$, or in the
  axiom \axiom{\enum{S}} specifying the \emph{partition} of an
  enumerated set $S$ to indicate the process classes concerned by the
  elements of the set; the latter applies for the axiom
  \axiom{\enum{\messPrefix}} in our example.  }

\begin{definition}[Local constants]
%%---------------------------------------------------------------------------
  Given a process $\proc\in \PCl$ and a constant $cst$, we say
  that $cst$ is \emph{local} to $\proc$ when  it is
  a function whose evaluation depends on $\proc$ ({\ie} of type  
  $\PCl\tfun cstType$ or $\NODES \tfun{} cstType$) or
  when it is (an element of an
  enumerated set whose partition is) annotated by $@\PCl$. 
  %
%% LONG version
\begin{longv}
  \textLV{By abuse of language we might sometimes say that $cst$ is local to
    $\PCl$.}
\end{longv}
%% LONG version
  % 
  We denote $\Cloc{\PCl}$ the set of local constants for (the
  processes of) $\PCl$.  
\end{definition}
%%---------------------------------------------------------------------------

In our example, the elements of $\messPrefix$ are local to both $p$
and $q\in Q$, $\neighbours(r)$ is local to any $r\in P\cup Q$ and
$availableResources(q)$ is local to any $q\in Q$.

\subsection{Producing Local Algorithms as State Machines}
\label{sec:machine}
%%============================================================================================
We specify now the algorithms for the set of processes.  Recall that
all processes in a process class run the same algorithm, the one
associated to the class.

%% \paragraph{Variables}
%% %
The machine \textsc{CM} in Figure~\ref{fig:global} declares the types
and initializes the local variables of each process class of the
distributed 
%% SHORT version
\begin{shortv}
  algorithm.
\end{shortv}
%% SHORT version
%% LONG version
\begin{longv}
  \textLV{algorithm; we use the same naming conventions for the
    corresponding axioms as for the constants in the context.}
\end{longv}
%% LONG version
The variable
$pc$ identifying the current state of each local algorithm
and the communication variable $\chanvar$ of type $\channel$ are defined
for any algorithm, the definition of other variables depends on the
modeled algorithm.
The variables together with their initialization
%% LONG version
\begin{longv}
  \textLV{and the corresponding invariants}
\end{longv}
%% LONG version
in the machine\textsc{ CM} modelling the algorithm
described in Example~\ref{ex:algoStar} is given in
Figure~\ref{fig:exampleMachine}.
%
%% LONG version
\begin{longv}
\begin{figure}[!t]
%% \framebox{\makebox[0.9\width]{
%% \resizebox{\textwidth}{!}{\boxed{

\centering 
\begin{minipage}{0.9\linewidth}
  \MACHINE{CM}{}{CONTEXT-CM}{}
\VARIABLES{
  \Variable{$\chanvar$ $pc$ $result$ $requestFrom$}{}
}
\INVARIANTS{
	\Invariant{channels\_typing}{false}{$\chanvar \in{} \channel$}{}
	\Invariant{pc\_typing}{false}{$pc\in{}\NODES\tfun{}STATES$}{}
	\Invariant{result\_typing}{false}{$result\in{}P\tfun{}(\NODES\pfun{}\nat{})$}{}
	\Invariant{requestFrom\_typing}{false}{$requestFrom\in Q\tfun \NODES$}{}
	\Invariant{channels\_respect\_network}{false}{$\forall x,y,m\cdot(x\in\NODES\land y\in\NODES\land m\in\MESSAGES\land \sent(\chanvar\mapsto(x\mapsto y)\mapsto m)>0\limp y\in \neighbours(x))$}{}
	\Invariant{requestFrom\_correctness}{false}{$\forall q\cdot(q\in Q\land pc(q)=\done \limp requestFrom(q)=\{p\})$}{}
	\Invariant{partial\_correctness}{false}{$pc(p) = done \limp{} result(p) = availableResources$}{}
}
\EVENTS{
\INITIALISATION{false}{}{
	\ACTIONS{false}{
		\Action{act1}{$\chanvar \bcmeq{} \emptychannel$}{true}{}
		\Action{act2}{$pc \bcmeq{} \{\proc\qdot{}\proc\in{}P|\proc\mapsto{}\sendingRequests\}\bunion{}\{q\qdot{}q\in{}Q|q\mapsto{}\waitingRequests\}$}{true}{}
		\Action{act3}{$result \bcmeq{} \{\proc\qdot{}\proc\in{}P|\proc\mapsto{}\emptyset{}\}$}{true}{}
		\Action{act4}{$requestFrom\bcmeq\{q\qdot q\in Q|q\mapsto\emptyset\}$}{true}{}
	}
}
%\vdots
}
\END
    
  \end{minipage}

\caption{Variables, invariants and initialisation for the Example~\ref{ex:algoStar}}
\label{fig:exampleMachine}
\end{figure}
\end{longv}
%% LONG version
%
%% SHORT version
\begin{shortv}
\begin{figure}[!t]
%% \framebox{\makebox[0.9\width]{
%% \resizebox{\textwidth}{!}{\boxed{
\centering 
\begin{minipage}{0.9\linewidth}
  \MACHINE{CM}{}{CONTEXT-CM}{}
\VARIABLES{
	%% \Variable{$\chanvar$ $pc$ $result$ $requestFrom$}{}
	\Variable{$\chanvar$ $pc$ $result$}{}
}
\INVARIANTS{
	\Invariant{channels\_typing}{false}{$\chanvar \in{} \channel$}{}
	\Invariant{pc\_typing}{false}{$pc\in{}\NODES\tfun{}STATES$}{}
	\Invariant{result\_typing}{false}{$result\in{}P\tfun{}(\NODES\pfun{}\nat{})$}{}
	%% \Invariant{requestFrom\_typing}{false}{$requestFrom\in Q\tfun \NODES$}{}
	%% %% \Invariant{channels\_respect\_network}{false}{$\forall x,y,m\cdot(x\in\NODES\land y\in\NODES\land m\in\MESSAGES\land \sent(\chanvar\mapsto(x\mapsto y)\mapsto m)>0\limp y\in \neighbours(x))$}{}
	%% %% \Invariant{requestFrom\_correctness}{false}{$\forall q\cdot(q\in Q\land pc(q)=\done \limp requestFrom(q)=\{p\})$}{}
	%% \Invariant{partial\_correctness}{false}{$pc(p) = done \limp{} result(p) = availableResources$}{}
}
\EVENTS{
\INITIALISATION{false}{}{
	\ACTIONS{false}{
		\Action{act1}{$\chanvar \bcmeq{} \emptychannel$}{true}{}
		\Action{act2}{$pc \bcmeq{} \{\proc\qdot{}\proc\in{}P|\proc\mapsto{}\sendingRequests\}\bunion{}\{\proc\qdot{}\proc\in{}Q|\proc\mapsto{}\waitingRequests{}\}$}{true}{}
		\Action{act3}{$result \bcmeq{} \{\proc\qdot{}\proc\in{}P|\proc\mapsto{}\emptyset{}\}$}{true}{}
		%% \Action{act4}{$requestFrom\bcmeq\{q\qdot q\in Q|q\mapsto\emptyset\}$}{true}{}
	}
}
%\vdots
}
\END
\end{minipage}
\caption{Variables, invariants and initialisation for the Example~\ref{ex:algoStar}}
\label{fig:exampleMachine}
\end{figure}
\end{shortv}
%% SHORT version

\begin{definition}[Local variables] % for a set of processes]
%%---------------------------------------------------------------------------
  Given a process $\proc\in \PCl$ and a variable $var$, we
  say that
  %% $var\in Var$
  $var$
  is \emph{local} to $\proc$ when it is a function
  whose evaluation depends on $\proc$ ({\ie} of type $\PCl\tfun
  varType$ or $\NODES \tfun{} varType$).
  %
  %% LONG version
  \begin{longv}
    \textLV{
      By abuse of language we might sometimes say that $var$ is local to
      $\PCl$.}
  \end{longv}
  %% LONG version
  %
  We denote
  %% $\Vloc{\PCl}\subseteq Var$
  $\Vloc{\PCl}$
  the set of local variables for (the
  processes of) $\PCl$ and $\Vloc{\proc}=\Vloc{\PCl}$ the set of local
  variables for a process $\proc\in\PCl$.
\end{definition}

%% LONG version
\begin{longv}
  \textLV{Every variable $var$ (except $\chanvar$) is local to one (or all)
    class(es) of processes as specified by a typing invariant $var\in
    \PCl\tfun varType$ (or $var\in\NODES\tfun varType$).}
\end{longv}
%% LONG version
Every variable is initialised as usual by a deterministic assignment
which specifies the value of the variable for the processes of each
concerned class using statements of the form $\{\proc\cdot
\proc\in\PCl|\proc\mapsto expr\}$ with the expression $expr$ using
\emph{only local constants and variables of the process $\proc$}.
%% LONG version
\begin{longv}
  \textLV{One can easily check that the initialisation expressions
    used in our example satisfy all the locality constraints.}
\end{longv}
%% LONG version
For example, the algorithm specific variable ${result}$ concerns only the
process $p\in{P}$ with the expression $result(p)$ corresponding to the
values received from the processes of $Q$.

%% LONG version
\begin{longv}
  \textLV{
    Note that a variable
    can be local to a process class or to all process classes; in the
    former category we have, in our example, the variable $result$ local
    to (processes of) $P$ and the variable $requestFrom$ local to
    (processes of) Q while in the latter we have the variable $pc$ which
    is local both to (processes of) $P$ and (processes of) $Q$.  The
    machine \textsc{CM} should contain a variable modelling
    communications, $\chanvar\in\channel$, which traces the state of
    communication channels.}
\end{longv}
%% LONG version

%% LONG version
\begin{longv}
  \textLV{Note that the invariant \axiom{partial\_correctness} expresses that,
if process $p$ terminates, then the result of the algorithm is
correct. This invariant was verified in our {\eventb} model.}
\end{longv}
%% LONG version

%% LONG version
\begin{longv}
  \textLV{ With respect to Tel's model
    (Section~\ref{sec:gener-defin-distr}) a local configuration is an
    element of the set $LS=Var\pfun Val$ and if we denote by
    $ls_{\proc}$ the local configuration of (the algorithm of) the
    process $\proc$, then the domain of $ls_{\proc}$ is exactly
    $\Vloc{\proc}$.  Moreover, the set of configurations of (the
    distributed algorithm of) the LB model is then ${\cal{C}} =
    (\NODES\tfun LS)\times \channel$.
    Note also that a configuration defines the values of all variables of
    an LB specification and is therefore equivalent to an \eventb{} state
    as defined in Section~\ref{sec:eevent-b-modelling}.}

  \textLV{Every local variable $var$ is defined in the clause
    \textsc{INVARIANTS} of the machine. In fact, $var(p)$ is the effective
    local variable of $p$ and it is sometimes written as $var_p$ (see for
    instance, G. Tel~\cite{tel2000introduction}).}
\end{longv}
%% LONG version
%%---------------------------------------------------------------------------

%% \paragraph{Events} % for a set of processes}
Events of the machine \textsc{CM} correspond to state transitions of
the local algorithms of the processes. Process events are observed
for a specific process of a process class.

%% \begin{definition}[Local  events, states of a process (class)]
\begin{definition}[LB  events, states]
%%---------------------------------------------------------------------------
  An \emph{event} \textsf{evt} in LB is such that
  \begin{itemize}
  \item it features one parameter $\proc$ typed by a guard $\proc\in
    \PCl$ with $\PCl\in\NODES$;
  \item all actions are
    %% deterministic
    assignments 
    $x(\proc) := pExpr$ or  $\chanvar := cExpr$ with $cExpr$ of the form
    \begin{itemize}
    \item $send(\chanvar\mapsto(\proc\mapsto pExpr)\mapsto mExpr)$
    \item $receive(\chanvar\mapsto(pExpr\mapsto \proc)\mapsto mExpr)$
    \end{itemize}
    If the event contains an action $\send$, resp. $\receive$, then it is
    called a \emph{send event}, resp \emph{receive event}; it is
    called \emph{internal} otherwise.
    
  \item it features a guard $pc(\proc)=st$ which specifies the event
    is enabled in state $st\in {\STATES}$;
  \item it features a typing guard $t\in tExpr$ for each parameter;
  \item if it is an internal or a send event, it can feature general guards $gExpr$ or guards of the form
    \begin{itemize}
    \item $\sent(\chanvar\mapsto(\proc\mapsto pExpr)\mapsto mExpr)=nExpr$
    \item $\received(\chanvar\mapsto(pExpr\mapsto \proc)\mapsto mExpr)=nExpr$
    %% \item    $\readyForReception(\chanvar\mapsto (pExpr\mapsto \proc)\mapsto mExpr)=bExpr$
    \end{itemize}
  \item if it is a receive event, it can feature matching guards for
    the parameters $source$ and $message$ which should be always
    present for such an event;
  \end{itemize}  
  with all expressions $tExpr,gExpr,pExpr,mExpr,nExpr$ built over
  local constants, local variables, parameters of the event, literal
  integers and booleans.

  We say that the event is observed for a process $\proc$ and moreover, that
  is \emph{observable} in state $st$.
  We denote by $\Evtloc{\PCl}$ and $\Evtloc{\proc}$ the set of local
  events for the set of processes $\PCl$ and for the process $\proc$
  respectively, and by $\Evtloc{\PCl,st}$ and $\Evtloc{\proc,st}$ the
  events in $\Evtloc{\PCl}$ and $\Evtloc{\proc}$ respectively, that
  are observable in state $st$. 

  Given a process class $\PCl$, the set of \emph{states of processes} of
  $\PCl$, denoted by $\States{\PCl}$, consists of the states $st$ such
  that there exists a parameter $\proc$ and a guard $pc(\proc)=st$ for
  some event $evt\in\Evtloc{\PCl}$.
\end{definition}
%%---------------------------------------------------------------------------

%% LONG version
\begin{longv}
  \textLV{
    %% We consider here a restriction of the above general definition of
    %% events to obtain a classification similar to the one proposed by
    %% Gerard Tel~\cite{tel2000introduction}.  The \event{send} and
    %% \event{receive} events are modifying the variable channel which is in
    %% fact a shared variable.  For each \event{send} or \event{receive}
    %% event only one message is sent or received.
    %% %
    %% We could add for convenience a \event{receive-and-send} event to allow
    %% more flexible models; one can split such an event into two separate
    %% \event{receive} and \event{send} events.
    The \event{send} and \event{receive} events are modifying the variable channel which is in
    fact a shared variable.
    Comparing to the classification  proposed by
    Gerard Tel~\cite{tel2000introduction} we also 
    add for convenience a \event{receive-and-send} event to allow
    more flexible models; one can split such an event into two separate
    \event{receive} and \event{send} events.
    The loss of a message can be
    modelled by an event modifying only the communication channels.}
%% \end{longv}
%% %% LONG version

%% %% LONG version
%% \begin{longv}
\textLV{ When considering a send event, one must verify that the
  destination of the message is a neighbour of the emitter.  As we can
  see in the invariants of our machine \textsc{CM} an invariant property
  \lbl{channels\_\allowbreak respect\_network} expresses that every
  sent messages has been sent to a neighbour of the sender.}
\end{longv}
%% LONG version

%
\begin{figure}[!t]
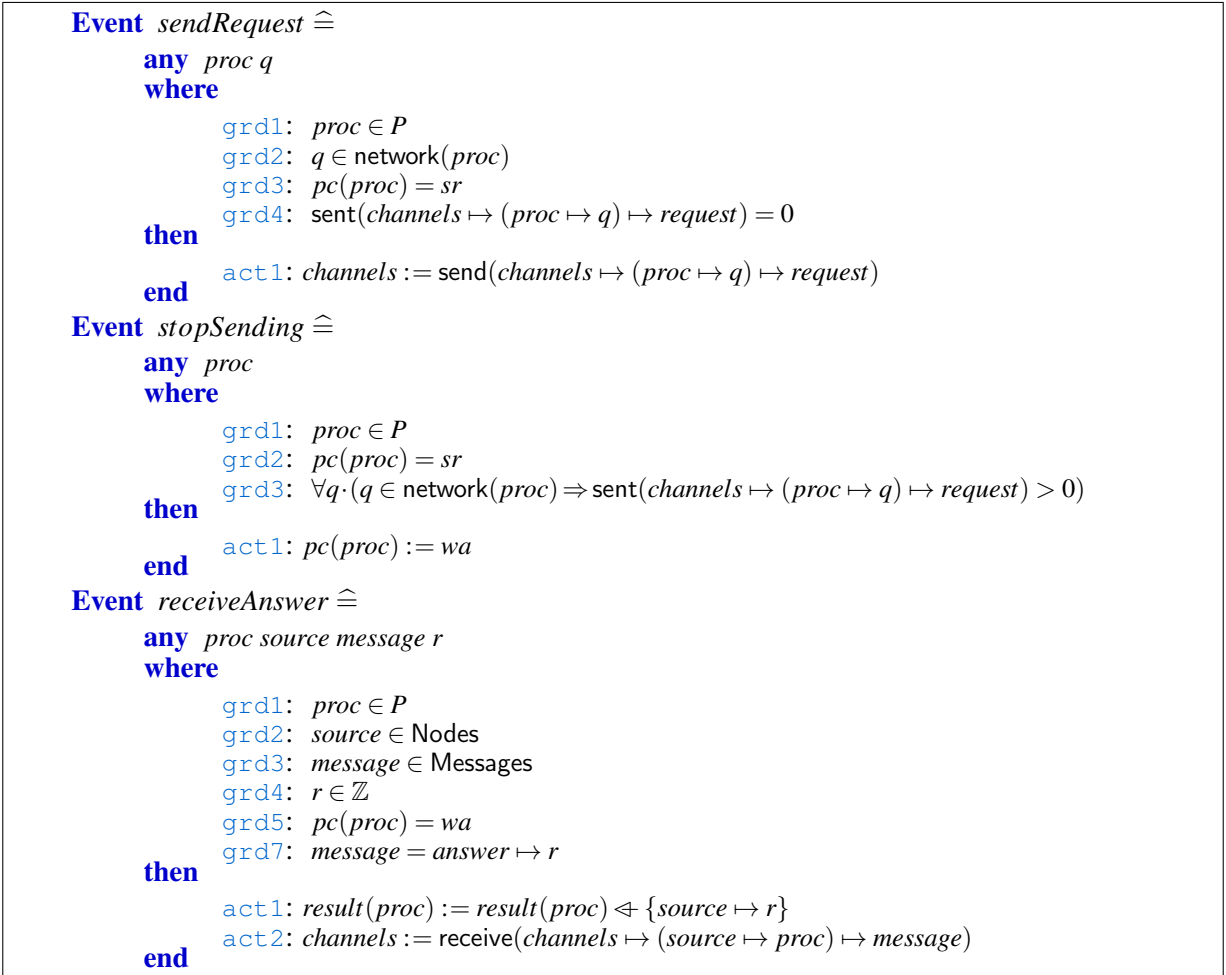

%% \framebox{\makebox[0.9\width]{
%% \resizebox{\textwidth}{!}{\boxed{
% \setstretch{0.8}
%
  \centering
  \begin{minipage}{0.9\linewidth}
\begin{description}
\MYEVT{\sendRequest{}}{false}{}{}{}{
	\ANY{
		\Param{$\proc$ $q$}{true}{}
	}
	\GUARDS{true}{
		\Guard{grd1}{false}{$\proc\in{}P$}{true}{}
		\Guard{grd2}{false}{$q\in{}\neighbours(\proc)$}{true}{}
		\Guard{grd3}{false}{$pc(\proc) = \sendingRequests$}{true}{}
		\Guard{grd4}{false}{$\sent(\chanvar\mapsto{}(\proc\mapsto{}q)\mapsto{}request)=0$}{true}{}
	}
	\ACTIONS{true}{
		\Action{act1}{$\chanvar \bcmeq{} \send(\chanvar\mapsto{}(\proc\mapsto{}q)\mapsto{}request)$}{true}{}
	}
}
\MYEVT{\stopSending{}}{false}{}{}{}{
	\ANY{
		\Param{$\proc$}{true}{}
	}
	\GUARDS{true}{
		\Guard{grd1}{false}{$\proc\in{}P$}{true}{}
		\Guard{grd2}{false}{$pc(\proc) = \sendingRequests$}{true}{}
		\Guard{grd3}{false}{$\forall{}q\qdot{}(q\in{}\neighbours(\proc)\limp{}\sent(\chanvar\mapsto{}(\proc\mapsto{}q)\mapsto{}request)>0)$}{true}{}
	}
	\ACTIONS{true}{
		\Action{act1}{$pc(\proc) \bcmeq{} \waitingAnswers$}{true}{}
	}
}
\MYEVT{\receiveAnswer{}}{false}{}{}{}{
	\ANY{
		\Param{$\proc$ $source$ $message$ $r$}{true}{}
	}
	\GUARDS{true}{
		\Guard{grd1}{false}{$\proc\in{}P$}{true}{}
		\Guard{grd2}{false}{$source\in{}\NODES$}{true}{}
		\Guard{grd3}{false}{$message\in{}\MESSAGES$}{true}{}
		\Guard{grd4}{false}{$r\in{}\intg{}$}{true}{}
		\Guard{grd5}{false}{$pc(\proc) = \waitingAnswers$}{true}{}
%%		\Guard{grd6}{false}{$\readyForReception(\chanvar\mapsto{}(source\mapsto{}\proc)\mapsto{}message)=TRUE$}{true}{}
		\Guard{grd7}{false}{$message = answer\mapsto{}r$}{true}{}
	}
	\ACTIONS{true}{
		\Action{act1}{$result(\proc) \bcmeq{} result(\proc)\ovl{}\{source\mapsto{}r\}$}{true}{}
		\Action{act2}{$\chanvar \bcmeq{} \receive(\chanvar\mapsto{}(source\mapsto{}\proc)\mapsto{}message)$}{true}{}
	}
}
\end{description}

  \end{minipage}

\caption{Events for the Example~\ref{ex:algoStar}}
\label{fig:exampleEvents}
\end{figure}

The events of the {\eventb} machine \textsc{CM} corresponding to the algorithm
for the process $p$ introduced in Example~\ref{ex:algoStar} are
presented in Figure~\ref{fig:exampleEvents}.
Note that $\sendRequest{}$ is a \event{send} event and does not modify $pc(p)$,
$\stopSending{}$ is an internal event with a guard verifying if $p$
has sent a request to all its neighbours, $\receiveAnswer{}$ is a receive
event for answers to the requests (the internal event $terminate$ not
presented here verifies that an answer has been received from every
neighbour and terminates the local algorithm of process $p$).
The processes of $Q$ feature similar events: we have a receive and a
send event which model respectively the reception of requests from $p$
and the dispatching of an
%% SHORT version
\begin{shortv}
  answer.
\end{shortv}
%% SHORT version
%% LONG version
\begin{longv}
  \textLV{answer with the stored value of the
concerned process.}
\end{longv}
%% LONG version
We also have an internal event
for terminating the local algorithm of a process of $Q$ once it has
sent the answer.
 %dm dm
\section{Translation in {\distAlgo}}
\label{sec:translation}
%%============================================================================================
A pair of a machine and a context compliant with the form described in
the previous section is translated towards a {\distAlgo} program
composed of a set of process 
%% SHORT version
\begin{shortv}
  classes.
\end{shortv}
%% SHORT version
%% LONG version
\begin{longv}
  \textLV{
  classes and a main function which defines
    the processes and starts them. Some specific additional restrictions on the context and machines are
    necessary for the translation into {\distAlgo}; we mention them
    explicitly only if not implicit in the presentation.
    In particular, the types of variables and constants cannot involve sets
    of  sets, sets of functions, functions on sets and functions on
    functions. Types cannot be relations in the current version.
    }
\end{longv}
%% LONG version
The main function and the process class definitions are generated from
the (axioms in the) context while the process class methods are
generated from the (invariants and events in the) machine.
%
%% LONG version
\begin{longv}
  \textLV{
  The generated main function purpose is to simulate the
    {\eventb} model on a single machine; in order to instantiate
    the processes on different machines some
    %% manual modifications
    external configuration might be needed.
    }
\end{longv}
%% LONG version

\subsection{Translation of Expressions}
\label{sec:translationExpr}
%%============================================================================================
We first define a translation function, denoted $\tr[\XX]{}$, which transforms a
well-formed {\eventb} expression (or predicate) $expr$ into the corresponding
{\distAlgo} code $\tr[\XX]{expr}$ {\wrt} a set $\XX$ of bound variables.

%% SHORT version
\begin{shortv}
  Arithmetic expressions are translated in an obvious way. Set
  expressions are also translated straightforwardly with sets built
  using the {\python} primitives
  {\lstinline[language=DistAlgo]?set?}  and
  {\lstinline[language=DistAlgo]?setof?}, and the set operations
  encoded by corresponding {\python} operations.  Finite functions are
  translated using {\python} dictionaries. Predicates are translated
  into boolean expressions with the quantifiers encoded using the
  \lstinline[language=DistAlgo]?each?  and
  \lstinline[language=DistAlgo]?some? {\distAlgo} functions. 
\end{shortv}
%% SHORT version

%% LONG version
\begin{longv}
  \textLV{
  Arithmetic expressions are translated in an obvious way, the only
worth mentioning case being that of intervals:}
\begin{equation*}
  %% \begin{aligned}
  \begin{array}{r@{\hspace{3pt}}c@{\hspace{3pt}}l@{\hspace{3pt}}}
    \tr[\XX]{e_1\upto e_2} &\eqdef& \text{\lstinline[language=DistAlgo]?set(range($\tr[\XX]{e_1}$,$\tr[\XX]{e_2}$+1))?}
  \end{array}
  %% \end{aligned}
  \end{equation*}
  
Set expressions are also translated in a straightforward way with
the empty set encoded by the {\python} expression {\lstinline[language=DistAlgo]?set()?}, a set
$\{e_1, \dots, e_n\}$ encoded by
{\lstinline[language=DistAlgo]?{$\tr[\XX]{e_1}$, ..., $\tr[\XX]{e_n}$}?} and the
set operations encoded by corresponding {\python} operations. In
particular coercions are translated as follows:
\begin{equation*}
  %% \begin{aligned}
  \begin{array}{r@{\hspace{3pt}}c@{\hspace{3pt}}l@{\hspace{3pt}}}
    \tr[\YY]{\{x_1,\dots,x_n\qdot x_1\in s_1\land\dots \land\ x_n\in s_n \land\ pred|expr\}}
    & \eqdef &
    \text{\lstinline[language=DistAlgo]?setof$(\tr[\YY\bun\XX]{expr}, x_1\ $
      in\ $\tr[\YY]{s_1},\dots,$ ?}\\
    &&
    \multicolumn{1}{r}{\text{\lstinline[language=DistAlgo]?$x_n\ $ in\ $\tr[\YY]{s_n}$, $\tr[\YY\bun\XX]{pred})$?} }
  \end{array}
%% \end{aligned}
\end{equation*}
with $\XX=\{x_1,\dots,x_n\}$ and $\tr[\XX]{x_i} = x_i$.

  Finite functions are translated using {\python} dictionaries which map keys to values:
\begin{equation*}
  %% \begin{aligned}
  \begin{array}{r@{\hspace{3pt}}c@{\hspace{3pt}}r@{\hspace{3pt}}}
    \tr[\XX]{\{e_1\mapsto v_1, \dots, e_n\mapsto v_n\}}
    &\eqdef&
    \text{\lstinline[language=DistAlgo]?${\tr[\XX]{e_1}$:$\tr[\XX]{v_1}$,$\ldots$,$\tr[\XX]{e_n}$:$\tr[\XX]{v_n}}$?}
  \end{array}
%% \end{aligned}
\end{equation*}
Besides the obvious translations for the classical operations on functions we have:
\begin{equation*}
  %% \begin{aligned}
  \begin{array}{r@{\hspace{3pt}}c@{\hspace{3pt}}l@{\hspace{3pt}}}
    \tr[\YY]{\{x_1,\dots,x_n\qdot x_1\in s_1\land\ \dots\land\ x_n\in s_n
      & \eqdef &
      \text{\lstinline[language=DistAlgo]?\{$\tr[\YY]{args}$:$\tr[\YY\bun\XX]{expr}\ $ for\ $x_1\ $ in\ $\tr[\YY]{s_1}$?}
      \\
      \multicolumn{1}{r}{\land\ pred|args\mapsto expr\}}}
    &&
    \multicolumn{1}{r}{\text{\lstinline[language=DistAlgo]?for\ $\ldots\ $ for\ $x_n\ $ in\ $\tr[\YY]{s_n}\ $ if\ $\tr[\YY\bun\XX]{pred}$\}?}}
    \\
    \tr[\YY]{\lambda x_1\mapsto\dots\mapsto x_n\qdot x_1\in s_1 \land\ \dots\land\ x_n \in s_n 
      & \eqdef &
      \text{\lstinline[language=DistAlgo]? \{($x_1$,$\dots$,$x_n$):$\tr[\YY\bun\XX]{expr}\ $for\ $x_1\ $ in\ $\tr[\YY]{s_1}$?}
      \\
      \multicolumn{1}{r}{\land\ pred | expr} }
    &&
    \multicolumn{1}{r}{\text{\lstinline[language=DistAlgo]? for\ $\ldots\ $ for\ $x_n\ $ in\ $\tr[\YY]{s_n}\ $ if\ $\tr[\YY\bun\XX]{pred}$\}?}}
    \\
    \tr[\XX]{f_1\bcmeq f_2}
    & \eqdef &
    \text{\lstinline[language=DistAlgo]?$\tr[\XX]{f_1}$=deepcopy($\tr[\XX]{f_2}$)?}
    \\
    \tr[\XX]{f_1\bcmeq f_1\ovl f_2}
    & \eqdef & \text{\lstinline[language=DistAlgo]?$\tr[\XX]{f_1}$.update(deepcopy($\tr[\XX]{f_2}$))?}
    \\
    \tr[\XX]{f(expr)}
    & \eqdef &
    \text{\lstinline[language=DistAlgo]?$\tr[\XX]{f}$[$\tr[\XX]{expr}$]?}
  \end{array}
  %% \end{aligned}
\end{equation*}
%% with $\XX=\{x_1, \dots, x_n\}$ and $\tr[\XX]{x_i}~=~x_i$}

  Predicates are translated into boolean expressions with a special care 
given to quantified variables:
\begin{equation*}
  %% \begin{aligned}
  \begin{array}{r@{\hspace{3pt}}c@{\hspace{3pt}}r@{\hspace{3pt}}}
    \tr[\YY]{\forall x_1, \dots, x_n\cdot(x_1\in s_1 \land\ \dots \land\ x_n\in s_n\limp pred)}
      & \eqdef &
      \text{\lstinline[language=DistAlgo]
        ?each(${x_1}\ $ in\ $\tr[\YY]{s_1}$,$\ldots$,${x_n}\ $in\ $\tr[\YY]{s_n}$?},\\
      %% \land\ x_n\in s_n\limp pred)}
    &&
    \text{\lstinline[language=DistAlgo]?has=$\tr[\YY\bun\XX]{pred}$)?}
    \\
    \tr[\YY]{\exists x_1, \dots, x_n\cdot(x_1\in s_1 \land\ \dots \land\ x_n\in s_n\land\ pred)} 
      & \eqdef &
      \text{\lstinline[language=DistAlgo]
        ?some(${x_1}\ $ in\ $\tr[\YY]{s_1}$,$\ldots$,${x_n}\ $in\ $\tr[\YY]{s_n}$?},
      \\
      %% \land\ x_n\in s_n\land\ pred)} 
    &&
    \text{\lstinline[language=DistAlgo]?has=$\tr[\YY\bun\XX]{pred}$)?}\\
  \end{array}
  %% \end{aligned}
\end{equation*}
\end{longv}
%% LONG version

The action for the sending of a message is translated using the
{\distAlgo} function \lstinline[language=DistAlgo]?send?:
\begin{equation*}
  %% \begin{aligned}
  \begin{array}{r@{\hspace{3pt}}c@{\hspace{3pt}}r@{\hspace{3pt}}}
    \tr[\XX]{\chanvar\bcmeq \send(\chanvar\mapsto(\proc\mapsto dest)\mapsto msg)}
      & \eqdef &
    \text{\lstinline[language=DistAlgo]?send($\tr[\XX]{msg}$, to=$\tr[\XX]{dest}$)?}
  \end{array}
  %% \end{aligned}
\end{equation*}
Note that $\chanvar$ and $\proc$ are not present in the resulting code
since $\chanvar$ is implicit in {\distAlgo} and $\proc$ corresponds to
the process
executing the \lstinline[language=DistAlgo]?send? statement.

The {\sent} and {\received} events  defined in {\eventb} are
translated as {\distAlgo} queries on message history.  {\distAlgo}
allows patterns inside queries on messages and any plain variable
\lstinline[language=DistAlgo]?x?  in such a query is considered free
and is potentially instantiated by a value following a successful
matching. To indicate that a variable is bound in a query it should be
of the form \lstinline[language=DistAlgo]?_x?.
We consider thus the translation  function $\trx[\XX]{}$ which is
defined exactly as the function $\tr[\XX]{}$ except for variables for
which we have
\begin{equation*}
  %% \begin{aligned}
  \begin{array}{r@{\hspace{3pt}}c@{\hspace{3pt}}l@{\hspace{30pt}}l}
    \trx[\overrightarrow{x}]{x}
    &\eqdef&
    \text{\lstinline[language=DistAlgo]?_x?}
    & \text{when } x\in \overrightarrow{x}
    \\
    \trx[\overrightarrow{x}]{x}
    &\eqdef&
    \tr[\XX]{x}
    & \text{when } x\not\in \overrightarrow{x}
  \end{array}
  %% \end{aligned}
\end{equation*}
The two expressions involving $sent$ or $received$ events
supported by our approach are translated using
%% LONG version
\begin{longv}
  \textLV{the
    \lstinline[language=DistAlgo]?sent? and
    \lstinline[language=DistAlgo]?received?  {\distAlgo} primitives:}
\end{longv}
%% LONG version
\begin{equation*}
  %% \begin{aligned}
  %   \begin{array}{l@{\hspace{3pt}}}
  \begin{array}{l@{\hspace{3pt}}c@{\hspace{2pt}}l@{\hspace{3pt}}}  
    \tr[\overrightarrow{x}]{sent(\chanvar\mapsto(\proc\mapsto dest)\mapsto msg)>0} 
     \eqdef 
    \text{\lstinline[language=DistAlgo]
      ?some(sent($\trx[\overrightarrow{x}]{msg}$,to=$\trx[\overrightarrow{x}]{dest}$))?}
    \\
    \tr[\overrightarrow{x}]{received(\chanvar\mapsto(source\mapsto \proc)\mapsto msg)>0}
       \eqdef 
      \text{\lstinline[language=DistAlgo]
        ?some(received($\trx[\overrightarrow{x}]{msg}$,?}
      \\
      %% \mapsto msg)>0}
          \multicolumn{1}{r}{\text{\lstinline[language=DistAlgo]
      ?from_=$\trx[\overrightarrow{x}]{source}$))?}}
  \end{array}
  %% \end{aligned}
\end{equation*}
%
%% LONG version
\begin{longv}
  \textLV{An equality test ($=0$) is translated by a negation of the form
    \lstinline[language=DistAlgo]?not(some(...))?.}

\textLV{For example, the expression 
$\forall{}q\qdot{}(q\in{}\neighbours{}(\proc)\limp{} sent(\chanvar\mapsto{}(\proc\mapsto{}q)\mapsto{}request)>0)$
from event $\stopSending$ defined in the previous section is translated into
\begin{equation*}
  %% \begin{aligned}
  \begin{tabular}{l@{\hspace{3pt}}l@{\hspace{3pt}}l@{\hspace{3pt}}}
    \lstinline[language=DistAlgo]
      ?each(q in $\tr[\emptyset]{\neighbours{}(\proc)}$,?
      \\
      \quad\quad\quad \lstinline[language=DistAlgo]
      ?has=some(sent(msg=($\tr[\emptyset]{request}$),to=_q)))?
  \end{tabular}
  %% \end{aligned}
\end{equation*}
with
$\tr[\emptyset]{\neighbours{}(\proc)}=$\lstinline[language=DistAlgo]?self?\neighbours{}
and
%% $\tr[\emptyset]{request}=$\lstinline[language=DistAlgo]?\messPrefix.request?
$\tr[\emptyset]{request}=$\lstinline[language=DistAlgo]?MessagePrefixes.request?
as explained in the next sections.}
\end{longv}
%% LONG version

~\\[-35pt]
\subsection{Generation of the Main Function}
\label{sec:transcont}
%%============================================================================================
The main function of the generated {\distAlgo} program defines
different local constants as well as the different processes to
execute, and starts the local algorithms of all the processes.  This
function is generated using exclusively the context
\textsc{CONTEXT-CM} and more precisely, only the axioms of the
context. The (identifiers of these) axioms should thus respect the
rules given in Section~\ref{sec:context} and the names of the
variables and constants are inferred correspondingly.

The code of the main function contains a fixed part independent of the
algorithm and specifying, for example, the behaviour of the
communication channels. We omit here the fixed part and the various
imports that might be needed and focus on the part generated from the
{\eventb} model.
%% \footnote{the complete version of the program is given
  %% in Appendix~\ref{app:program}.}

The axiom \axiom{\NODESax}
%% :$partition(\NODES, \PCl_1,\dots,\PCl_n)$
allows us to infer the set $\{\PCl_1,\dots,\PCl_n\}$ of process
classes and to generate, for each process class, a fresh variable
\lstinline[language=DistAlgo]?PClSet$_i$?  corresponding to the set of
processes in $\PCl_i$.
We can thus initialize each variable
\lstinline[language=DistAlgo]?PClSet$_i$? as a set of
\lstinline[language=DistAlgo]?NPCl$_i$? processes of class
\lstinline[language=DistAlgo]?PCl$_i$? (generated later on) and then,
the variable \lstinline[language=DistAlgo]?Nodes? corresponding to
the set of all processes:
\begin{lstlisting}[language=DistAlgo]
    PClSet$_1$ = new(PCl$_1$, num=NPCl$_1$)
    $\ldots$
    PClSet$_n$ = new(PCl$_n$, num=NPCl$_n$)
    Nodes = set.union(PClSet$_1$,$\ldots$,PClSet$_n$)
\end{lstlisting}
We use the axioms $\axiom{\PCl}_i$ 
%% :$partition(\PCl, \{\proc_1\},\dots,\{\proc_n\})$
to initialize the variables for each set and
\lstinline[language=DistAlgo]?NPCl$_i$? to the cardinal of the
corresponding set (\lstinline[language=DistAlgo]?NPCl$_i$? should be configured
manually if the axiom is not present):
\begin{lstlisting}[language=DistAlgo]
    (~\proc~1,$\ldots$,~\proc~m) = list(PClSet$_i$)
    NPCl$_i$ = $|\{\proc_1,\dots,\proc_m\}|$
\end{lstlisting}
 
Starting from the axiom \axiom{network\_value}
we generate the map
%% \neighbours{}
\lstinline[language=DistAlgo]?network? for the
topology
\begin{lstlisting}[language=DistAlgo]
    network = {~\proc~:$\tr[\emptyset]{expr_1}$ for ~\proc~ in PClSet$_1$}
    network.update({~\proc~:$\tr[\emptyset]{expr_2}$ for ~\proc~ in PClSet$_2$})
    $\ldots$
    network.update({~\proc~:$\tr[\emptyset]{expr_n}$ for ~\proc~ in PClSet$_n$})
\end{lstlisting}
%
%% LONG version
\begin{longv}
  \textLV{If this axiom is not present in the {\eventb} context, then it should
    be filled by hand in {\distAlgo}}.
\end{longv}
%% LONG version

In fact, for each (local) constant $cst$ in the context which is a
function ($cst \in \PCl \tfun cstType$) and features an axiom
\axiom{cst\_value}: $cst=\{\proc\cdot \proc \in \PCl|\proc\mapsto
expr\}$ for some $\PCl$ we generate an initialization:
\begin{lstlisting}[language=DistAlgo]
    cst = {~\proc~:$\tr[\emptyset]{expr}$ for ~\proc~ in PClSet}
\end{lstlisting}

For each process class $\PCl_i$ the following code is
generated for the initialisation:
\begin{lstlisting}[language=DistAlgo]
    for ~\proc~ in PClSet$_i$:
        setup({~\proc~}, (cst_1[~\proc~],$\ldots$,cst_n[~\proc~])
\end{lstlisting}
with $\{cst_1,\ldots,cst_n\}=\Cloc{\PCl_i}$.

Finally, the processes are executed with the {\distAlgo} command
\lstinline[language=DistAlgo]?start(Nodes)?.

\begin{example}
\label{ex:main}
%%---------------------------------------------------------------------------
Given the context in Section~\ref{sec:context} the following main
function is generated.
%, basicstyle=\scriptsize]
\begin{lstlisting}[language=DistAlgo]
    def main():
        NP = 1
        NQ = #NQ - to be configured
    
        PSet = new(P, num=NP)
        (p,) = list(PSet)
        QSet = new(Q, num=NQ)
    
        Nodes  = set.union(PSet, QSet)
        network = {~\proc~:QSet for ~\proc~ in PSet}
        network.update({q:{p} for q in QSet})
        availableResources = #availableResources - to be configured
        
        for ~\proc~ in PSet:
            setup({~\proc~}, (network[~\proc~],))
        for ~\proc~ in QSet:
            setup({~\proc~}, (network[~\proc~], availableResources[~\proc~]))
        start(Nodes) 
\end{lstlisting}
%%---------------------------------------------------------------------------
\end{example}

In the same time with the main class we generate the code
corresponding to the enumerated sets defined in the context using an
axiom $\axiom{\enum{S}}:{partition(S,\{el_1\},\{el_2\},\dots)}$
like, {\eg}, $\messPrefix$.
For all these sets
%%except for {\NODES}, process classes, {\STATES} and {\MESSAGES}, 
we generate a separate file (imported when needed) containing the
corresponding
%% {\distAlgo}
code:
\begin{lstlisting}[language=DistAlgo]
        class S(Enum):
            $el_1$ = "$el_1$"
            $el_2$ = "$el_2$"
            ...
\end{lstlisting}
The access to the elements of the respective set is done as expected:
$\tr[\XX]{el_i}$~\eqdef~\lstinline[language=DistAlgo]?S.$el_i$?, for
any member $el_i$ of the enumerated set.

\subsection{Generation of the Process Classes}
\label{sec:transmach}
%%============================================================================================
For each process class $\PCl$ we generate
%% LONG version
\begin{longv}
  \textLV{
    (in an individual file)}
\end{longv}
%% LONG version
a {\distAlgo} process class \lstinline[language=DistAlgo]?PCl?
featuring the necessary methods.
%% LONG version
\begin{longv}
  \textLV{More precisely, we generate
    %% which correspond to the local algorithm of this class:
    the \lstinline[language=DistAlgo]?setup? method, the
    \lstinline[language=DistAlgo]?run? method,
    \lstinline[language=DistAlgo]?receive? methods, and additional methods
    for the events concerned by the process class.}
\end{longv}
%% LONG version

For the purpose of the translations
presented
in this section we
consider the function $\trp[\XX]{}$ which behaves exactly like
$\tr[\XX]{}$ except for one case:
$\trp[\XX]{f(\proc)}\eqdef$\lstinline[language=DistAlgo]?self.f? when
$f\in\Vloc{\PCl}\bunion\Cloc{\PCl}$, $\proc\in\PCl$.
%% for some process class $\PCl$.

%% \commentHC{not really the only exception - for lcal vars/csts as well
  %% as said below.}

The \lstinline[language=DistAlgo]?setup? method gets the values of the
local constants as parameters and initializes the local variables. We
have thus for each process class $\PCl$ in the context a {\distAlgo}
class:
\begin{lstlisting}[language=DistAlgo]
    class PCl( process ) :
        def setup($cst_1,\ldots,cst_n$):
            self.$var_1 = \trp[\emptyset]{expr_1}$
            $\vdots$
            self.$var_m = \trp[\emptyset]{expr_m}$
\end{lstlisting}
with $\{cst_1,\ldots,cst_n\}=\Cloc{\PCl}$,
$\{var_1,\ldots,var_m\}=\Vloc{\PCl}$, and
$\{expr_1,\ldots,expr_m\}$ the corresponding  
expressions 
$var_i \bcmeq \{\proc\cdot \proc\in\PCl|\proc\mapsto expr_i\}$ 
in the {\textcolor{keycolor}{\textbf{Initialisation}}} section of the
machine.
%% (Figure~\ref{fig:exampleMachine}).
For a variable $var$ (resp. constant $cst$), the translation of
$var(\proc)$ (resp. $cst(\proc)$) is then
\lstinline[language=DistAlgo]?self.var?
(resp. \lstinline[language=DistAlgo]?self.cst?).

For each state $st\in \States{\PCl}$ a method
\lstinline[language=DistAlgo]?st? describing the behavior on reception
of an event observable in state $st$ is generated as explained below.
The \lstinline[language=DistAlgo]?run? method defining the control flow
of the program for the respective process consists of a loop which
calls at each iteration the method \lstinline[language=DistAlgo]?st?
corresponding to the current value of
\lstinline[language=DistAlgo]?self.pc? and terminates when
\lstinline[language=DistAlgo]?self.pc? reaches the termination state
{\done}.
When $\States{\PCl} = \{st_1,\dots,st_n\}$  the following code is
generated:
\begin{lstlisting}[language=DistAlgo]
    def run():
        stateFunctions = {"st1":st1,$\dots$,"stn":stn}
        while(self.pc!=done):
        stateFunctions[self.pc]()
\end{lstlisting}~\\[-20pt]

Given an event $evt\in\Evtloc{\PCl}$ we denote by $\grds{evt}$ the set
of its guards, by $\acts{evt}$ the set of its actions and by
$\Params{evt}$ the set of its parameters.  The translation $\tri{}$ of
a set of guards of an \event{internal} or a \event{send} event is as
follows:
\begin{equation*}
  %% \begin{aligned}
  \begin{array}{r@{\hspace{3pt}}c@{\hspace{3pt}}l@{\hspace{3pt}}}
    \tri{\{\proc\in\PCl, t_1\in S_1,\ldots, t_l\in S_l,
      &\eqdef&
    \text{\lstinline[language=DistAlgo]?self.pc=="st" and some($t_1$
      in $\trp[\emptyset]{S_1}$, $\dots$, $t_l$ in $\trp[\emptyset]{S_l}$,?}
    \\
    pc(\proc)=st, g_1,\ldots, g_n\} } &&
    \multicolumn{1}{r}{\text{\lstinline[language=DistAlgo]?has=$\trp[\Params{evt}]{g_1}$ and $\ldots$ and $\trp[\Params{evt}]{g_n}$)?}}
  \end{array}
  %% \end{aligned}
\end{equation*}
where $\Params{evt}=\{t_1,\ldots,t_l\}$ and $S_1,\ldots,S_l$ are
finite sets.  The translation $\tra[\XX]{\acts{evt}}$ of a set of
actions of an \event{internal} or \event{send} event $evt$ is defined
as the juxtaposition of the translations $\trp[\XX]{a_j}$ of each
action in the set $\acts{evt}$.  
Since the actions of $\acts{evt}$
are observed concurrently but translated as a sequence of
assignments, fresh temporary variables are defined as copies of the
local variables prior to the event and are used to access the old
values of the local variables. However, for simplicity, we omit
these temporary fresh variables
%% SHORT version
\begin{shortv}
  in our example.
\end{shortv}
%% SHORT version
%% LONG version
\begin{longv}
  \textLV{in this section and explicit them
  only in the appendix.}
\end{longv}
%% LONG version

For each state $st\in \States{\PCl}$ we use the set
$\{evt_1,\dots,evt_m\}\subseteq\Evtloc{\PCl,st}$
of all \event{internal} and \event{send} events observable in state
$st$ to generate the method \lstinline[language=DistAlgo]?st?:
\begin{lstlisting}[language=DistAlgo]
    def st():
        --st
        if await ($\tri{\grds{evt_1}}$):
            $\tra[\Params{evt_1}]{\acts{evt_1}}$
        $\vdots$
        elif($\tri{\grds{evt_m}}$):
            $\tra[\Params{evt_m}]{\acts{evt_m}}$
        elif(self.pc != "st"):
            pass
\end{lstlisting}~\\[-10pt]
with the label \lstinline[language=DistAlgo]?--st? and the keyword
\lstinline[language=DistAlgo]?await? added only if there is a
\event{receive} event in $\Evtloc{\PCl,st}$; this statement is used to
enable the reception of messages.
When an \lstinline[language=DistAlgo]?await?  statement is reached
every message that has arrived to destination but has not been processed
yet, \ie{} messages in the message queue of this process,  is handled
(using the
\lstinline[language=DistAlgo]?receive? methods) before the
\lstinline[language=DistAlgo]?if?  conditions are evaluated.
Messages are received until the message queue is empty and one of the guard conditions is satisfied.

\begin{example}
\label{ex:sending}
%%---------------------------------------------------------------------------
In our example, we have $\Evtloc{P, \sendingRequests}= \{\sendRequest{},
\stopSending{}\}$ and thus the following code is generated for the
method \lstinline[language=DistAlgo]?sr?.\\

\begin{lstlisting}[language=DistAlgo]
    def sr():
        # event ~\sendRequest~
        if(self.pc == "sr" and
            some(q in self.~\neighbours{}~,
                has=not(some(sent((~\messPrefix~.request,), to=_q))))):
            send((~\messPrefix~.request,), to=q)
        # event ~\stopSending~
        elif(self.pc == "sr" and
              each(q in self.~\neighbours{}~,
                  has=some(sent((~\messPrefix~.request,), to=_q)))):
            self.pc = "wa"
        elif(self.pc != "sr"):
            pass
\end{lstlisting}
%%---------------------------------------------------------------------------
\end{example}

For each \event{receive} event $evt$ in $\Evtloc{\PCl,st}$ we generate
a
\lstinline[language=DistAlgo]?receive? method in the class
\lstinline[language=DistAlgo]?PCl?:
\begin{lstlisting}[language=DistAlgo]
    def receive($\trr{\grds{evt}}$):
        $\tra[\Params{evt}]{\acts{evt}}$
\end{lstlisting}
where the translation $\trr{\grds{evt}}$ of a set of guards of a
\event{receive} event $evt$ is as follows:
\begin{equation*}
  %% \begin{aligned}
  \begin{array}{r@{\hspace{3pt}}c@{\hspace{3pt}}l@{\hspace{3pt}}}
    \trr{\{\proc\in\PCl, msg\in \MESSAGES, source\in \NODES, t_1\in S_1,\ldots, t_l\in S_l,
      &\eqdef&
      \text{\lstinline[language=DistAlgo]?msg=($\tr[\emptyset]{msgExpr}$),?}
      \\
      pc(\proc)=st, msg=msgExpr, source=procExpr)\} }
    &&
    \text{\lstinline[language=DistAlgo]?from_=$\tr[\emptyset]{procExpr}$,?}
    \\
    &&
      \text{\lstinline[language=DistAlgo]?at=(st,)?}
  \end{array}
  %% \end{aligned}
\end{equation*}

If $procExpr$ is empty, {\ie} not specified in the model then
  a free variable it is used in the translation (to indicate the
  source of the message is not specified). We proceed similarly when
  $msgExpr$ is empty.
%
%% LONG version
\begin{longv}
  \textLV{The parameters $\Params{evt}$ are used in the expressions
    $msgExpr$ and $procExpr$ to specify the expected message $msg$
    that is received from the emitter $source$.  Note that the
    reception of the message is performed automatically by {\distAlgo}
    before the \lstinline[language=DistAlgo]?receive? method is
    executed and therefore the reception action in LB does not need to
    be translated; only the message handler should be translated.}
\end{longv}
%% LONG version
%
The actions of a \event{receive} event are translated in the same way
as the actions of an \event{internal} or \event{send} event.

%% LONG version
\begin{longv}
  \textLV{
    The \lstinline[language=DistAlgo]?receive?  methods of {\distAlgo}
    ({\ie} the message handlers) also have the particularity that if
    multiple \lstinline[language=DistAlgo]?receive?  methods can be
    executed for the reception of a message, they will all be executed.
    Therefore, the \event{receive} events of the model must have exclusive
    guards in order to have only one
    \lstinline[language=DistAlgo]?receive? method executed at a time.}
\end{longv}
%% LONG version
  
\begin{example}
The following code corresponds to the \event{receive} event $\receiveAnswer{}$.
%, basicstyle=\scriptsize]
\begin{lstlisting}[language=DistAlgo]
    def receive(msg=(~\messPrefix~.answer, r), from_=source,
                at=(wa,)):
        self.result[source] = r
\end{lstlisting}
\end{example}
  
{The translation has been implemented in \textsf{Java} as a {\rodin}
plugin and the source code together with the installation instructions
are available at \url{https://gitlab.inria.fr/agrall/eb2da}.
}
 %dm
\section{Concluding Remarks and Future Work}
\label{sec:concl-remar-future}
%%============================================================================================
%%We only consider a static topology in this paper; for dynamic one the topology should be specified as a variable.
%% All these communication operations may be introduced in a
%% theory\cite{rodinTheory} using the {\rodin} theory plugin.
 
The localization of {\eventb} has been used when a distributed
algorithm \cite{ieee1394,abrial2010} has been developed using the
correct-by-construction paradigm and especially the refinement
relationship among levels of abstractions. The translation of local
{\eventb} models was a manual process and the current work provides a
systematic way to produce a {\distAlgo} program from a local {\eventb}
model.

{We claim the LB modelling language is sufficiently powerful
  to model a large variety of distributed algorithms and abstract
  enough to be considered as the basis for the translation towards
  different target distributed programming languages. A couple of algorithms
  have been modelled and the programs obtained by translation allowed
  the simulation of the algorithms for different numbers of nodes.  We
  continue to develop more and more elaborated case studies.}

{In the short term we plan of course to produce the proof of
  soundness of the translation.
  The communication model used for the algorithms implemented so far
  although reliable does not guarantee the order of messages; we 
  intend to provide the model for other communications models
  together with the corresponding translation.
  At the implementation level, we should first provide an automatic
  packaging and facilitate the installation as a {\rodin} plugin.
  The definition of
  transformations for other target distributed programming languages is
  a more long term objective.
}

 %dm

%% \newpage

\bibliographystyle{eptcs}
\bibliography{biblio}
%\end{document}
\newpage
\begin{longv}
\input{appendix.tex}
\end{longv}
\end{document}